\documentclass[10pt]{article}

\usepackage[utf8]{inputenc}
\usepackage{tikz}
\usetikzlibrary{shapes,arrows}
\usepackage[none]{hyphenat} % para no cortar las palabras
\usepackage{amsmath}
\usepackage{amssymb}
\usepackage{amsfonts}
\usepackage{multirow}
\usepackage{natbib}
\usepackage{enumerate}
\usepackage{hyperref}
\usepackage{graphicx}
\hypersetup{colorlinks=true, linkcolor=blue, urlcolor=blue, citecolor=red}
\usepackage{listings}
\usepackage{color}
\usepackage{epstopdf}
\usepackage{rotating}
\usepackage{algorithm}
\usepackage{algorithmic}

\usepackage[top=2cm, bottom=2cm, left=2cm, right=2cm]{geometry}

\title{\bf  Generalized Ridge Regression: Applications to Nonorthogonal Linear Regression Models}
\author{
    \bf{Rom\'an Salmer\'on G\'omez}\thanks{Professor, Department of Quantitative methods for economics and business, University of Granada, Spain (e-mail: romansg@ugr.es).}
    \and
    \bf{Catalina Garc\'ia Garc\'ia}\thanks{Professor, Department of Quantitative methods for economics and business, University of Granada, Spain (e-mail: cbgarcia@ugr.es).}
    \and 
    \bf{Guillermo Hortal Reina}\thanks{PhD student. University of Granada. Spain  (e-mail: ghorrei@correo.ugr.es).}
}
\date{\today}

\begin{document}

\maketitle

\maketitle

\begin{abstract}
This paper analyzes the possibilities of using the generalized ridge regression to mitigate multicollinearity in a multiple linear regression model. For this purpose, we obtain the expressions for the estimated variance, the coefficient of variation, the coefficient of correlation, the variance inflation factor and the condition number.
The results obtained are illustrated with two numerical examples. 
\end{abstract}

\section{Introduction}

Take the following multiple linear regression model with $n$ observations and $p$ variables (intercept included):
\begin{equation}\label{modelo}
    \mathbf{Y} = \mathbf{X} \boldsymbol{\beta} + \mathbf{u},
\end{equation}
where $E[\mathbf{u}] = \mathbf{0}$ and $var \left( \mathbf{u} \right) = \sigma^{2} \mathbf{I}$, and the goal is to estimate the unknown parameters $\boldsymbol{\beta}$ for the establishment of relations between the $p$ independent variables in the matrix $\mathbf{X}$ and the dependent variable $\mathbf{Y}$. The well-known ordinary least squares (OLS) estimator is usually applied: $\widehat{\boldsymbol{\beta}} = \left( \mathbf{X}^{t} \mathbf{X} \right)^{-1} \mathbf{X}^{t} \mathbf{Y}$.

Estimation using the OLS requires that the matrix $\mathbf{X}$ presents a rank equal to $p$. Thus, it is assured that the columns of matrix $\mathbf{X}$ are linearly independent; and consequently, the inverse of matrix $\mathbf{X}^{t} \mathbf{X}$ exists. However, even if the rank condition is verified, an inexact but high linear relation between these columns can lead to problems in the OLS estimation that can provide inconclusive
numerical and statistical analysis. For this reason, we consider that it is relevant to detect if the existing linear relations can lead to these problems and what methodologies can be applied to mitigate this situation.

The ridge regression is one of the most applied methodologies as an alternative to OLS and was initially presented by \cite{HoerlKennard1970a,HoerlKennard1970b}. Both papers are focused on the regular case in which a positive constant $k$ is introduced in the main diagonal of matrix $\mathbf{X}^{t} \mathbf{X}$, although the generalized ridge regression is also briefly presented. Recently, \cite{hoerl2020ridge} and \cite{hastie2020ridge} have reviewed the origins of ridge regression and its developments and extensions within the field of applied statistics.

It is well known that the regular ridge (RR) estimator is given by the following expression:
\begin{equation}\label{ridge_k}
\widehat{\boldsymbol{\beta}}(k) = \left( \mathbf{X}^{t} \mathbf{X} + k \mathbf{I} \right)^{-1} \mathbf{X}^{t} \mathbf{Y}, \quad k \geq 0.
\end{equation}
This improves the conditioning of matrix $\mathbf{X}^{t} \mathbf{X}$, and the results show that matrix $\mathbf{X}^{t} \mathbf{X} + k \mathbf{I}$ is more stable than $\mathbf{X}^{t} \mathbf{X}$. Although the properties of this estimator are widely analyzed in the works of Hoerl and Kennard, to the best of our knowledge, there are no similar references for the case of the generalized estimator other than the work recently presented by \cite{SalmeronGRR2021}.

Following \cite{SalmeronGRR2021}, the generalized ridge (GR) estimator is given by the following expression:
\begin{equation}\label{ridge_K}
\widehat{\boldsymbol{\beta}}(\mathbf{K}) = \left( \mathbf{X}^{t} \mathbf{X} + \boldsymbol{\Gamma} \mathbf{K} \boldsymbol{\Gamma}^{t} \right)^{-1} \mathbf{X}^{t} \mathbf{Y},
\end{equation}
where $\boldsymbol{\Gamma}$ is an orthonormal matrix that contains the eigenvectors $\mathbf{X}^{t} \mathbf{X}$, and $\boldsymbol{\Lambda}$ is a diagonal matrix that contains its eigenvalues, which are real and positive. In addition, the matrix $\mathbf{K}$ is a diagonal matrix whose elements $k_{i}$, $i=1,\dots,p$, are real and nonnegative.

\cite{SalmeronGRR2021} showed that this estimator is biased. The augmented model leads to this estimator and its matrix of variances and covariances, its norm, its mean squared error and a measure for the goodness of fit being obtained. All these measures are analyzed for the generalized case where $k_{i} \geq 0$, for the regular case where ($\mathbf{K} = k \mathbf{I}$) and for the particular case of the generalized case in which $\mathbf{K} = diag(0,\dots,k_{l},\dots,0)$ where $k_{l} \geq 0$ for $l=1,\dots,p$. It is shown that this last case presents better behavior than the regular case, especially in relation to the mean squared error.
It is also proposed to perform inference on the coefficients of the independent variables of the model using bootstrap methodology (see, for example, Efron \cite{Efron1979,Efron1981,Efron1982,Efron1987}, Efron and Gong \cite{EfronGong1983} or the a review of the above-mentioned works presented by Efron and Tibshirani \cite{EfronTibshirani1986}).

Now, in this second work, we analyze the potential of the generalized ridge regression to mitigate the multicollinearity existing in a model similar to the one given in (\ref{modelo}), paying special attention to the particular case in which $\mathbf{K} = diag(0,\dots,k_{l},\dots,0)$. Thus, Section \ref{ev} analyzes the matrix of variances and covariances of the GR estimator, Section \ref{CV} analyzes its coefficient of variation (CV), Section (\ref{CC}) analyses its coefficient of correlation and Sections \ref{VIF} and \ref{cn} propose expressions for the variance inflation factor (VIF) and the condition number (CN), respectively. Prior to this, Section \ref{ndt} reviews the consequences of data transformation when the ridge regression is applied and summarizes the notations that will be applied throughout the paper. Finally, Section \ref{example} illustrates the contribution of this paper with two numerical examples, and Section \ref{conclusion} summarizes the main conclusions.

\section{Data Transformation and Notation}\label{ndt}

\subsection{Data Transformation}

\cite{Garciaetal2016}, \cite{Salmeron2018CN} and \cite{Rodriguez2019R2,Rodriguezetal2021} presented extensions of the VIF, CN, goodness of fit and the Stewart index (see \cite{Stewart1987}) to the ridge regression, concluding that \textit{the transformation of the data is not optional but is required for the correct application of these measures in ridge regression}.
This question was also addressed by Marquardt in 1980 in the paper entitled ``You should standardize the predictor variables in your regression models'' (\cite{Marquardt1980}).

Therefore, in this paper, the data will be standardized. Thus, if $\mathbf{x}$ and $\mathbf{y}$ are the standardized matrices of $\mathbf{X}$ and $\mathbf{Y}$, respectively, it is verified that $\mathbf{x}^{t} \mathbf{x}$ contains the correlations of the variables included in $\mathbf{X}$ and $\mathbf{x}^{t} \mathbf{y}$ contains the correlations of the variables included in $\mathbf{X}$ with $\mathbf{Y}$.

To standardize a vector, each of its elements is subtracted from its mean and divided by the square root of $n$ times its variance:
\begin{itemize}
\item If $\mathbf{X} = \left( X_{ij} \right)_{i=1,\dots,n; j=1\dots,p}$, then $\mathbf{x} = \left( x_{ij} \right)_{i=1,\dots,n; j=2,\dots,p}$, where $x_{ij} = \frac{X_{ij} - \overline{\mathbf{X}}_{j}}{\sqrt{n var \left( \mathbf{X}_{j} \right)}}$ with $\overline{\mathbf{X}}_{j}$ and $var \left( \mathbf{X}_{j} \right)$ being the mean and the variance of the variable $\mathbf{X}_{j}$, $j=2,\dots,p$, respectively. In this case, $\overline{\mathbf{x}}_{j} = 0$ and $var \left( \mathbf{x}_{j} \right) = \frac{1}{n}$, where $\mathbf{x}_{j}$ is column $j$ of $\mathbf{x}$. Furthermore, $\mathbf{R} = \mathbf{x}^{t} \mathbf{x}$ is the matrix of correlations of the variables included in $\mathbf{x}$.
\item If $\mathbf{Y} = \left( y_{i} \right)_{i=1,\dots,n}$, then $\mathbf{y} = \left( y_{i} \right)_{i=1,\dots,n}$, where $y_{i} = \frac{Y_{i} - \overline{\mathbf{Y}}}{\sqrt{n var \left( \mathbf{Y} \right)}}$ with $\overline{\mathbf{Y}}$ and $var \left( \mathbf{Y} \right)$ being the mean and the variance of variable $\mathbf{Y}$, respectively. In this case, $\overline{\mathbf{y}} = 0$ and $var \left( \mathbf{y} \right) = \frac{1}{n}$.
\end{itemize}
Due to this transformation, the intercept disappears in the new model, i.e., a new noncentered model is obtained with $m = p-1$ independent variables.

\subsection{Notation}

The notation used in the work and the expressions obtained in the work of \cite{SalmeronGRR2021} are summarized below:
\begin{description}
\item[Model:] $\mathbf{y} = \mathbf{x} \boldsymbol{\beta} + \mathbf{u}$, where $\mathbf{x}_{n \times m}$, $\mathbf{y}_{n \times 1}$, $\boldsymbol{\beta}_{m \times 1}$, $\mathbf{u}_{n \times 1}$, $n$ is the number observations and $m$ is the number of independent variables (without intercept).
\item[OLS estimator:] $\widehat{\boldsymbol{\beta}} = \left( \mathbf{x}^{t} \mathbf{x} \right)^{-1} \mathbf{x}^{t} \mathbf{y}$.
\item[Decomposition of eigenvalues and eigenvectors:] $\mathbf{x}^{t} \mathbf{x} = \boldsymbol{\Gamma} \boldsymbol{\Lambda} \boldsymbol{\Gamma}^{t}$, where $\boldsymbol{\Lambda} = diag (\lambda_{1},\dots, \lambda_{m})$, $\lambda_{1},\dots, \lambda_{m} \in \mathbb{R}^{+}$ and $\boldsymbol{\Gamma} \boldsymbol{\Gamma}^{t} = \mathbf{I} = \boldsymbol{\Gamma}^{t} \boldsymbol{\Gamma}$ with $\mathbf{I}$ an identity matrix of appropriate dimensions in each case. Thus, if $\boldsymbol{\Gamma} = \left( \gamma_{ij} \right)_{i,j=1,\dots,m}$, then $\sum \limits_{j=1}^{m} \gamma_{ij}^{2} = 1 = \sum \limits_{j=1}^{m} \gamma_{ji}^{2}$ for $i=1,\dots,m$ and $\sum \limits_{j=1}^{m} \gamma_{ij} \gamma_{hj} = 0 = \sum \limits_{j=1}^{m} \gamma_{ji} \gamma_{jh}$ for $i,h = 1,\dots,m$ and $i \not= h$.
\item[Generalized ridge estimator:] $\widehat{\boldsymbol{\beta}}(\mathbf{K}) = \left( \mathbf{x}^{t} \mathbf{x} + \boldsymbol{\Gamma} \mathbf{K} \boldsymbol{\Gamma}^{t} \right)^{-1} \mathbf{x}^{t} \mathbf{y} = \boldsymbol{\Gamma} \boldsymbol{\Omega} \boldsymbol{\delta}$ where $\mathbf{K} = diag ( k_{1},\dots, k_{m} )$ with $k_{i} \geq 0$ for $i = 1,\dots,m$, $\boldsymbol{\Omega} = \left( \boldsymbol{\Lambda} + \mathbf{K} \right)^{-1} = diag \left( \frac{1}{\lambda_{1}+k_{1}}, \dots, \frac{1}{\lambda_{m}+k_{m}} \right)$ and $\boldsymbol{\delta} = \boldsymbol{\Gamma}^{t} \boldsymbol{\alpha}$ with $\boldsymbol{\alpha} = \mathbf{x}^{t} \mathbf{y}$.
\item[Matrices of variances and covariances] $var \left( \widehat{\boldsymbol{\beta}} \right) = \sigma^{2}  \boldsymbol{\Gamma}  \boldsymbol{\Lambda}^{-1}  \boldsymbol{\Gamma}^{t}$ and        $var \left( \widehat{\boldsymbol{\beta}}(\mathbf{K}) \right) = \sigma^{2}  \boldsymbol{\Gamma}  \boldsymbol{\Psi}  \boldsymbol{\Gamma}^{t}$, where $\boldsymbol{\Psi} = \boldsymbol{\Omega} \boldsymbol{\Lambda} \boldsymbol{\Omega} = diag \left( \frac{\lambda_{1}}{(\lambda_{1}+k_{1})^{2}}, \dots, \frac{\lambda_{m}}{(\lambda_{m}+k_{m})^{2}} \right)$.
\item[Norm:] $|| \widehat{\boldsymbol{\beta}}(\mathbf{K}) || = \boldsymbol{\delta}^{t}  \boldsymbol{\Omega}^{2}  \boldsymbol{\delta}$.
\item[Mean Squared Error:] $MSE \left( \widehat{\boldsymbol{\beta}}(\mathbf{K}) \right) = \sigma^{2}  \boldsymbol{\Gamma}  \boldsymbol{\Psi}  \boldsymbol{\Gamma}^{t} + \boldsymbol{\xi}^{t}  \boldsymbol{\Theta}  \boldsymbol{\xi}$, where $\boldsymbol{\xi} = \boldsymbol{\Gamma}^{t} \boldsymbol{\beta}$ and $\boldsymbol{\Theta} = \left( \boldsymbol{\Omega}  \boldsymbol{\Lambda} - \mathbf{I} \right)  \left( \boldsymbol{\Omega}  \boldsymbol{\Lambda} - \mathbf{I} \right) = diag \left( \frac{k_{1}^{2}}{(\lambda_{1}+k_{1})^{2}}, \dots, \frac{k_{m}^{2}}{(\lambda_{m}+k_{m})^{2}} \right)$.
\item[Goodness of Fit:] $GoF (\mathbf{K}) = \frac{\widehat{\boldsymbol{\beta}}(\mathbf{K})^{t}  \left( \mathbf{x}^{t} \mathbf{x} + 2 \boldsymbol{\Gamma} \mathbf{K} \boldsymbol{\Gamma}^{t} \right) \widehat{\boldsymbol{\beta}}(\mathbf{K})}{\mathbf{y}^{t} \mathbf{y}} = \widehat{\boldsymbol{\beta}}(\mathbf{K})^{t}  \left( \mathbf{x}^{t} \mathbf{x} + 2 \boldsymbol{\Gamma} \mathbf{K} \boldsymbol{\Gamma}^{t} \right) \widehat{\boldsymbol{\beta}}(\mathbf{K})$ since $\mathbf{y}^{t} \mathbf{y} = 1$.
\item[Augmented model:] $\mathbf{y}_{a} = \mathbf{x}_{a} \boldsymbol{\beta} + \mathbf{u}_{a}$ where $\mathbf{y}_{a} = \left(\begin{array}{c}
                \mathbf{y} \\
                \mathbf{0}
            \end{array} \right)_{(n+m) \times 1}$,
 being $\mathbf{0}$ a vector composed of zeros with dimension $m \times 1$ and $\mathbf{x}_{a} = \left(
            \begin{array}{c}
                \mathbf{x} \\
                \mathbf{K}^{1/2}  \boldsymbol{\Gamma}^{t}
            \end{array} \right)_{(n+m) \times m}$ verifying that $\widehat{\boldsymbol{\beta}}_{a} = \left( \mathbf{x}^{t}_{a} \mathbf{x}_{a} \right)^{-1} \mathbf{x}^{t}_{a} \mathbf{y}_{a} = \widehat{\boldsymbol{\beta}}(\mathbf{K})$. Taking into account that:
            $$\mathbf{K} = \left(
                \begin{array}{ccccc}
                    k_{1} & \dots & 0 & \dots & 0 \\
                    \vdots &  & \vdots &  & \vdots \\
                    0 & \dots & k_{l} & \dots & 0 \\
                    \vdots &  & \vdots &  & \vdots \\
                    0 & \dots & 0 & \dots & k_{m}
                \end{array} \right), \quad \boldsymbol{\Gamma} = \left(
                \begin{array}{ccccc}
                    \gamma_{11} & \dots & \gamma_{1l} & \dots & \gamma_{1m} \\
                    \vdots &  & \vdots &  & \vdots \\
                    \gamma_{l1} & \dots & \gamma_{ll} & \dots & \gamma_{lm} \\
                    \vdots &  & \vdots &  & \vdots \\
                    \gamma_{m1} & \dots & \gamma_{ml} & \dots & \gamma_{mm}
                \end{array} \right),$$
            the following is obtained:
            \begin{equation}
                \mathbf{x}_{a} = \left(
                \begin{array}{ccccc}
                    \mathbf{x}_{12} & \dots & \mathbf{x}_{1l} & \dots & \mathbf{x}_{1m} \\
                    \vdots &  & \vdots &  & \vdots \\
                    \mathbf{x}_{l2} & \dots & \mathbf{x}_{ll} & \dots & \mathbf{x}_{lm} \\
                    \vdots &  & \vdots &  & \vdots \\
                    \mathbf{x}_{n2} & \dots & \mathbf{x}_{nl} & \dots & \mathbf{x}_{nm} \\
                    \sqrt{k_{1}} \gamma_{11} & \dots & \sqrt{k_{1}} \gamma_{l1} & \dots & \sqrt{k_{1}} \gamma_{m1} \\
                    \vdots &  & \vdots &  & \vdots \\
                    \sqrt{k_{l}} \gamma_{1l} & \dots & \sqrt{k_{l}} \gamma_{ll} & \dots & \sqrt{k_{l}} \gamma_{ml} \\
                    \vdots &  & \vdots &  & \vdots \\
                    \sqrt{k_{m}} \gamma_{1m} & \dots & \sqrt{k_{m}} \gamma_{lm} & \dots & \sqrt{k_{m}} \gamma_{mm}
                \end{array} \right).
                \label{x_aumenta}
            \end{equation}
    \end{description}

\section{Estimated Variance}
    \label{ev}

One of the consequences that may appear when the linear relation in a multiple linear regression is troubling is that the variances of the OLS estimators will be larger than those that will be obtained if the variables are orthogonal (see \cite{OBrien} for more details). For this reason, it is interesting to check if the matrix of variances-covariances of the generalized ridge regression is lower than that of the OLS since this fact will mitigate the undesirable effects of a troubling degree of multicollinearity.
Thus:
$$var \left( \widehat{\boldsymbol{\beta}}(\mathbf{K}) \right) - var \left( \widehat{\boldsymbol{\beta}} \right) = \sigma^{2} \boldsymbol{\Gamma} \left( \boldsymbol{\Psi} - \boldsymbol{\Lambda}^{-1} \right) \boldsymbol{\Gamma}^{t},$$
where:
$$\boldsymbol{\Psi} - \boldsymbol{\Lambda}^{-1} = diag \left( - \frac{k_{1}^{2} + 2 \lambda_{1} k_{1}}{(\lambda_{1}+k_{1})^{2} \lambda_{1}}, \dots, - \frac{k_{m}^{2} + 2 \lambda_{m} k_{m}}{(\lambda_{m}+k_{m})^{2} \lambda_{m}} \right).$$

Considering that $\mathbf{a} \in \mathbb{R}^{m}$ with $\mathbf{a} \not= \mathbf{0}$, it is verified that:
$$\mathbf{a}^{t} \boldsymbol{\Gamma} \left( \boldsymbol{\Psi} - \boldsymbol{\Lambda}^{-1} \right) \boldsymbol{\Gamma}^{t} \mathbf{a} = \mathbf{b}^{t} \left( \boldsymbol{\Psi} - \boldsymbol{\Lambda}^{-1} \right) \mathbf{b} = - \sum \limits_{j=1}^{m} \frac{k_{j}+2\lambda_{j}k_{j}}{(\lambda_{j}+k_{j})^{2} \lambda_{j}} b_{j}^{2} <0,$$
where $\mathbf{b} = \boldsymbol{\Gamma}^{t} \mathbf{a}$. Consequently, $\boldsymbol{\Gamma} \left( \boldsymbol{\Psi} - \boldsymbol{\Lambda}^{-1} \right) \boldsymbol{\Gamma}^{t}$ is a negative defined matrix; and since $\sigma^{2}>0$, it is verified that:
$$var \left( \widehat{\boldsymbol{\beta}}(\mathbf{K}) \right) - var \left( \widehat{\boldsymbol{\beta}} \right) <0 \rightarrow var \left( \widehat{\boldsymbol{\beta}}(\mathbf{K}) \right) < var \left( \widehat{\boldsymbol{\beta}} \right).$$

\section{Coefficient of Variation}
    \label{CV}

\cite{MarquardtSnee} and \cite{SneeMarquardt} distinguished between essential and nonessential multicollinearity. Thus, essential multicollinearity is the linear relation between the independent variables of a linear regression model similar to the one given by (\ref{modelo}) excluding the intercept, and nonessential multicollinearity is the linear relation between the intercept and at least one of the resting independent variables (with a very low variability, practically constant). \cite{Salmeron2020a} showed that nonessential multicollinearity can be detected according to a low CV. In addition, it is known that this problem is resolved by centering the independent variable (subtracting the mean gives a coefficient of variation equal to infinity).

Recently, \cite{Salmeron2020centered} provided another definition of the nonessential multicollinearity that generalized the concept given by Marquardt and Snee. It is understood that there is generalized nonessential multicollinearity when there are two independent variables with low variability that are related.
If one of the variables is the intercept, the definition given by Marquardt and Snee will be recovered.

Considering the expression given in (\ref{x_aumenta}) for $\mathbf{x}_{a}$ and considering that $\mathbf{x}_{a}^{t} \mathbf{x}_{a} = \mathbf{x}^{t} \mathbf{x} + \boldsymbol{\Gamma} \mathbf{K} \boldsymbol{\Gamma}^{t}$, the augmented model with standardized variables is a noncentered model, i.e., a model where there is no intercept. Thus, the nonessential multicollinearity defined by Marquardt and Snee cannot exist, but its generalized version was provided by \cite{Salmeron2020centered}.

Therefore, it could be useful to obtain an expression for the coefficient of variation in the ridge regression to determine the variability of each of the independent variables (columns) included in $\mathbf{x}_{a}$.

Considering that $\mathbf{x}_{a,l}$ is the $l$ columns of $\mathbf{x}_{a}$ given in (\ref{x_aumenta}), the following is obtained:
\begin{itemize}
\item Its mean is equal to $\overline{\mathbf{x}}_{a,l} = \frac{1}{n+m} \left( \sum \limits_{j=1}^{m} x_{jl} + \sum \limits_{j=1}^{m} \sqrt{k_{j}} \gamma_{lj} \right) = \frac{\sum \limits_{j=1}^{m} \sqrt{k_{j}} \gamma_{lj}}{n+m}$ since $\sum \limits_{j=1}^{m} x_{jl} =0$ as the data are standardised.
\item Its variance is equal to $var \left( \mathbf{x}_{a,l} \right) = \frac{1}{n+m} \left( \sum \limits_{j=1}^{m} x_{jl}^{2} + \sum \limits_{j=1}^{m} k_{j} \gamma_{lj}^{2} \right) - \overline{\mathbf{x}}_{a,l}^{2}$.
\end{itemize}
Thus, the coefficient of variation of $\mathbf{x}_{a,l}$ is given by the following expression:
$$CV \left( \mathbf{x}_{a,l}, \mathbf{K} \right) = \frac{\sqrt{var \left( \mathbf{x}_{a,l} \right)}}{| \overline{\mathbf{x}}_{a,l} |}, \quad l=1,\dots,m.$$

To analyze its monotony, its squares are obtained:
\begin{equation}\label{cv_general}
  CV \left( \mathbf{x}_{a,l}, \mathbf{K} \right)^{2} = \frac{var \left( \mathbf{x}_{a,l} \right)}{\overline{\mathbf{x}}_{a,l}^{2}} = \frac{\frac{1}{n+m} \left( \sum \limits_{j=1}^{m} x_{jl}^{2} + \sum \limits_{j=1}^{m} k_{j} \gamma_{lj}^{2} \right)}{\frac{1}{(n+m)^{2}} \left(  \sum \limits_{j=1}^{m} \sqrt{k_{j}} \gamma_{lj} \right)^{2}} - 1, \quad l=1,\dots,m.
\end{equation}

Note that when $k_{j}$ increases for all $j$s, the value of $CV \left( \mathbf{x}_{a,l}, \mathbf{K} \right)^{2}$ decreases. Thus, the degree of generalized nonessential multicollinearity increases.

\subsection{Particular cases}
\begin{itemize}
\item In the particular case when $\mathbf{K} = k \mathbf{I}$, i.e., $k_{1} = \dots = k_{m}=k$, the expression (\ref{cv_general}) can be rewritten as:
$$CV \left( \mathbf{x}_{a,l}, k \right)^{2} =  \frac{n+m}{k \left( \sum \limits_{j=1}^{m} \gamma_{lj} \right)^{2}} \sum \limits_{j=1}^{m} x_{jl}^{2} + n + m -1.$$
due to $\sum \limits_{j=1}^{m} \gamma_{jl}^{2} = 1$.

\item In the particular case when $\mathbf{K} = diag(0,\dots,k_{l},\dots,0)$, $l=1,\dots,m$, the expression (\ref{cv_general}) can be rewritten as:
$$CV \left( \mathbf{x}_{a,l}, k_{l} \right)^{2} =  (n+m) \left( \frac{\sum \limits_{j=1}^{m} x_{jl}^{2} + k_{l} \gamma_{ll}^{2}}{(\sqrt{k_{l}} \gamma_{ll})^{2}} \right) - 1 = \frac{n+m}{k_{l} \gamma_{ll}^{2}} \sum \limits_{j=1}^{m} x_{jl}^{2} + n + m -1.$$
\end{itemize}
In both cases, because the square root is an increasing monotonic function, considering the absolute value, it is obtained that the coefficient of variation decreases as a function of $k$, $k_{l}$ and is continuous in $k = 0 = k_{l}$ (due to its coincidence with the CV of $\mathbf{x}_{l}$). Consequently, as previously commented, the variability of $\mathbf{x}_{a,l}$ decreases as $k$ and $k_{l}$ increases. In other words, the degree of generalized nonessential multicollinearity increases. This fact is not surprising because we take part of the data from a set with standardized data where this type of multicollinearity does not exist.

Furthermore, since:
$$\lim \limits_{k \rightarrow + \infty} CV \left( \mathbf{x}_{a,l}, k \right) = \sqrt{n+m-1} = \lim \limits_{k_{l} \rightarrow + \infty} CV \left( \mathbf{x}_{a,l}, k_{l} \right),$$
the degree of multicollinearity will not be troubling. \cite{Salmeron2020a} stated that nonessential multicollinearity will be troubling when the CV is lower than 0.1002506.

\section{Coefficient of Correlation}
    \label{CC}

Given the variables $\mathbf{x}_{a,i}$ and $\mathbf{x}_{a,j}$ (columns $i$ and $j$ of matrix $\mathbf{x}_{a}$ given in (\ref{x_aumenta})), the coefficient of correlation is given by the following expression:
$$corr(\mathbf{x}_{a,i},\mathbf{x}_{a,j},\mathbf{K}) = \frac{cov(\mathbf{x}_{a,i},\mathbf{x}_{a,j},\mathbf{K})}{\sqrt{var(\mathbf{x}_{a,i},\mathbf{K})} \sqrt{var(\mathbf{x}_{a,j},\mathbf{K})}}, \quad i,j = 1,\dots,m, \ i \not=j,$$
where:
\begin{eqnarray*}
        cov(\mathbf{x}_{a,i},\mathbf{x}_{a,j},\mathbf{K}) &=& \frac{1}{n+m} \sum \limits_{h=1}^{n+m} x_{a,hi} x_{a,hj} - \overline{\mathbf{x}}_{a,i} \overline{\mathbf{x}}_{a,j} \\
            &=& \frac{1}{n+m} \left( \sum \limits_{h=1}^{n} \left( \frac{X_{hi} - \overline{\mathbf{X}}_{i}}{\sqrt{n var(\mathbf{X}_{i})}} \right) \left( \frac{X_{hj} - \overline{\mathbf{X}}_{j}}{\sqrt{n var(\mathbf{X}_{j})}} \right) + \sum \limits_{h=n+1}^{n+m} \sqrt{k_{h-n}} \gamma_{i h-n} \sqrt{k_{h-n}} \gamma_{j h-n}  \right) \\
            & & - \left( \frac{1}{n+m} \sum \limits_{l=1}^{m} \sqrt{k_{l}} \gamma_{il} \right) \left( \frac{1}{n+m} \sum \limits_{l=1}^{m} \sqrt{k_{l}} \gamma_{jl} \right) \\
            &=& \frac{1}{n+m} \left( corr(\mathbf{X}_{i}, \mathbf{X}_{j}) + \sum \limits_{l=1}^{m} k_{l} \gamma_{il} \gamma_{jl} \right) - \frac{1}{(n+m)^{2}} \left( \sum \limits_{l=1}^{m} \sqrt{k_{l}} \gamma_{il} \right) \left( \sum \limits_{l=1}^{m} \sqrt{k_{l}} \gamma_{jl} \right),
    \end{eqnarray*}
\begin{equation*}
        var(\mathbf{x}_{a,i},\mathbf{K}) = \frac{1}{n+m} \sum \limits_{h=1}^{n+m} x_{a,hi}^{2} - \overline{\mathbf{x}}_{a,i}= \\
           \frac{1}{n+m} \left( 1 + \sum \limits_{l=1}^{m} k_{l} \gamma_{il}^{2} \right) - \frac{1}{(n+m)^{2}} \left( \sum \limits_{l=1}^{m} \sqrt{k_{l}} \gamma_{il} \right)^{2}.
    \end{equation*}

\subsection{Particular cases}
\begin{itemize}
\item
In the particular case when $\mathbf{K} = k \mathbf{I}$, it is obtained that:
$$corr(\mathbf{x}_{a,i},\mathbf{x}_{a,j},k) = \frac{corr(\mathbf{X}_{i}, \mathbf{X}_{j}) - \frac{k}{n+m} \left( \sum \limits_{l=1}^{m} \gamma_{il} \right)\left( \sum \limits_{l=1}^{m} \gamma_{jl} \right)}{\sqrt{ 1 + k \left( 1 - \frac{1}{n+m} \left( \sum \limits_{l=1}^{m} \gamma_{il} \right)^{2} \right)} \sqrt{ 1 + k \left( 1 - \frac{1}{n+m} \left( \sum \limits_{l=1}^{m} \gamma_{jl} \right)^{2} \right)}},$$
due to $\sum \limits_{l=1}^{m} \gamma_{il} \gamma_{jl} = 0$ and $\sum \limits_{l=1}^{m} \gamma_{il}^{2} = 1$. Observing this expression, the following is concluded:
\begin{itemize}
\item It is a function continuous in $k$. I.e., for $k=0$, it is obtained that $corr(\mathbf{x}_{a,i},\mathbf{x}_{a,j},0) = corr(\mathbf{X}_{i}, \mathbf{X}_{j})$.
\item If $\left( \sum \limits_{l=1}^{m} \gamma_{il} \right)^{2}, \left( \sum \limits_{l=1}^{m} \gamma_{jl} \right)^{2} < n+m$ is a function that decreases as $k$ increases since the variance decreases when $k$ increases and the variances increase.
\item Its limit when $k \rightarrow +\infty$ is:
$$\frac{-\frac{1}{n+m} \left( \sum \limits_{l=1}^{m} \gamma_{il} \right)\left( \sum \limits_{l=1}^{m} \gamma_{jl} \right)}{\sqrt{1 - \frac{1}{n+m} \left( \sum \limits_{l=1}^{m} \gamma_{il} \right)^{2}} \sqrt{1 - \frac{1}{n+m} \left( \sum \limits_{l=1}^{m} \gamma_{jl} \right)^{2}}}.$$
This limit is higher than -1 whenever $\left( \sum \limits_{l=1}^{m} \gamma_{il} \right)^{2} + \left( \sum \limits_{l=1}^{m} \gamma_{jl} \right)^{2} < n+m$. In that case, considering the conditions of continuity and monotony, $corr(\mathbf{x}_{a,i},\mathbf{x}_{a,j}) \in [0, 1]$ for all values of $k$.
\end{itemize}

\item In the particular case when $\mathbf{K} = diag (0,\dots,k_{l},\dots,0)$, $l=1,\dots,m$, it is verified that:
$$corr(\mathbf{x}_{a,i},\mathbf{x}_{a,j},k_{l}) = \frac{corr(\mathbf{X}_{i}, \mathbf{X}_{j}) + \frac{n+m-1}{n+m} k_{l} \gamma_{il} \gamma_{jl}}{\sqrt{1 + \frac{n+m-1}{n+m} k_{l} \gamma_{il}^{2}} \sqrt{1 + \frac{n+m-1}{n+m} k_{l} \gamma_{jl}^{2}}}.$$
Observing this expression, the following is concluded:
\begin{itemize}
\item It is a function continuous in $k_{l}$, i.e., for  $k_{l}=0$, it is obtained that $corr(\mathbf{x}_{a,i},\mathbf{x}_{a,j},0) = corr(\mathbf{X}_{i}, \mathbf{X}_{j})$.
\item If $k_{l}$ increases, the variances and covariances also increase. Consequently, it is not possible to conclude a relation to the monotony of the correlation. Consequently, it is not assured that $corr(\mathbf{x}_{a,i},\mathbf{x}_{a,j})$ belongs to the interval $[0, 1]$ for all values of $k_{l}$.
\item Its limit when $k_{l} \rightarrow +\infty$ is:
$$\frac{\frac{n+m-1}{n+m} \gamma_{il} \gamma_{jl}}{\sqrt{\frac{n+m-1}{n+m} \gamma_{il}^{2}} \sqrt{\frac{n+m-1}{n+m} \gamma_{jl}^{2}}} = \pm 1.$$
Thus, there is perfect multicollinearity (the sign will depend on the values of $\gamma_{il}$ and $\gamma_{jl}$) between $\mathbf{x}_{a,i}$ and $\mathbf{x}_{a,j}$ for $k_{l} \rightarrow +\infty$. Consequently, the determinant of the matrix of correlation will tend to zero.
\end{itemize}
\end{itemize}
Then, it is possible to conclude that in the particular case when $\mathbf{K} = diag (0,\dots,k_{l},\dots,0)$, $l=1,\dots,m$, it is inappropriate to mitigate the essential multicollinearity. However, as its monotony is not defined, values of $k_{l}$ that mitigate it could exist.

\section{Variance Inflation Factor}\label{VIF}

The variance inflation factor (VIF) for coefficient $l$ of the ridge regression is obtained from the coefficient of determination of the auxiliary regression, where $\mathbf{x}_{a,l}$ is the dependent variable and the independent variables are the resting variables included in $\mathbf{x}_{a}$.
Because this regression is a noncentered regression (without an intercept), the measure of the goodness of fit presented by \cite{Salmeron2020centered} or \cite{SalmeronGRR2021}, which coincides with the coefficient of determination if the dependent variable presents zero mean, is considered.

In that case, the VIF is defined as:
$$VIF(l, \mathbf{K}) = \frac{1}{1 - R_{l,\mathbf{K}}^{2}}, \quad l=1,\dots,m,$$
where $R_{l,\mathbf{K}}^{2}$ is the coefficient of determination of the auxiliary regression.
\begin{equation}
        \label{reg_aux}
        \mathbf{x}_{a,l} = \mathbf{x}_{a,-l} \boldsymbol{\nu} + \boldsymbol{\epsilon},
    \end{equation}
where $\mathbf{x}_{a,-l}$ eliminates column $l$ of matrix $\mathbf{x}_{a}$.

Defining the residuals of model (\ref{reg_aux}) as $\mathbf{e}_{l} = \mathbf{x}_{a,l} - \mathbf{x}_{a,-l} \widehat{\boldsymbol{\nu}}$, it is obtained that $\mathbf{x}_{a,l} = \mathbf{x}_{a,-l} \widehat{\boldsymbol{\nu}} + \mathbf{e}_{l}$; furthermore, as a consequence, it is verified that
\begin{eqnarray*}
        \mathbf{x}_{a,l}^{t} \mathbf{x}_{a,l} &=& \left( \mathbf{x}_{a,-l} \widehat{\boldsymbol{\nu}} + \mathbf{e}_{l} \right)^{t} \left( \mathbf{x}_{a,-l} \widehat{\boldsymbol{\nu}} + \mathbf{e}_{l} \right) = \widehat{\boldsymbol{\nu}}^{t} \mathbf{x}_{a,-l}^{t} \mathbf{x}_{a,-l} \widehat{\boldsymbol{\nu}} + 2 \widehat{\boldsymbol{\nu}}^{t} \mathbf{x}_{a,-l}^{t} \mathbf{e}_{l} + \mathbf{e}_{l}^{t} \mathbf{e}_{l} \\
        &=& \widehat{\boldsymbol{\nu}}^{t} \mathbf{x}_{a,-l}^{t} \mathbf{x}_{a,-l} \widehat{\boldsymbol{\nu}} + \mathbf{e}_{l}^{t} \mathbf{e}_{l},
    \end{eqnarray*}
due to considering that $\widehat{\boldsymbol{\nu}} = \left( \mathbf{x}_{a,-l}^{t} \mathbf{x}_{a,-l} \right)^{-1} \mathbf{x}_{a,-l}^{t} \mathbf{x}_{a,l}$:
$$\widehat{\boldsymbol{\nu}}^{t} \mathbf{x}_{a,-l}^{t} \mathbf{e}_{l} = \widehat{\boldsymbol{\nu}}^{t} \mathbf{x}_{a,-l}^{t} \left( \mathbf{x}_{a,l} - \mathbf{x}_{a,-l} \widehat{\boldsymbol{\nu}} \right) = \widehat{\boldsymbol{\nu}}^{t} \left( \mathbf{x}_{a,-l}^{t} \mathbf{x}_{a,l} - \mathbf{x}_{a,-l}^{t} \mathbf{x}_{a,-l} \widehat{\boldsymbol{\nu}} \right) = \widehat{\boldsymbol{\nu}}^{t} \left( \mathbf{x}_{a,-l}^{t} \mathbf{x}_{a,l} - \mathbf{x}_{a,-l}^{t} \mathbf{x}_{a,l} \right) = 0.$$

In this case, it is obtained that:
$$R_{l,\mathbf{K}}^{2} = \frac{\mathbf{x}_{a,l}^{t} \mathbf{x}_{a,-l} \left( \mathbf{x}_{a,-l}^{t} \mathbf{x}_{a,-l} \right)^{-1} \mathbf{x}_{a,-l}^{t} \mathbf{x}_{a,l}}{\mathbf{x}_{a,l}^{t} \mathbf{x}_{a,l}}.$$
Then:
\begin{equation}
        VIF(l, \mathbf{K}) = \frac{\mathbf{x}_{a,l}^{t} \mathbf{x}_{a,l}}{\mathbf{x}_{a,l}^{t} \mathbf{x}_{a,l} - \mathbf{x}_{a,l}^{t} \mathbf{x}_{a,-l} \left( \mathbf{x}_{a,-l}^{t} \mathbf{x}_{a,-l} \right)^{-1} \mathbf{x}_{a,-l}^{t} \mathbf{x}_{a,l}}, \quad l=1,\dots,m. \label{VIF_K}
    \end{equation}

\subsection{Particular cases}
\begin{itemize}
\item
In the particular case when $\mathbf{K} = k \mathbf{I}$, $\mathbf{x}_{a,l}$ and $\mathbf{x}_{a,-l}$ can be expressed as:
$$\mathbf{x}_{a,l} = \left(
\begin{array}{c}
            \mathbf{x}_{l} \\
           \sqrt{k} \boldsymbol{\gamma}_{l}^{t}
        \end{array} \right), \quad
\mathbf{x}_{a,-l} = \left(
\begin{array}{c}
            \mathbf{x}_{-l} \\
            \sqrt{k} \boldsymbol{\Gamma}_{-l}^{t}
        \end{array} \right),$$
where $\boldsymbol{\gamma}_{l}$ is row $l$ of $\boldsymbol{\Gamma}$ and $\boldsymbol{\Gamma}_{-l}$ is the results obtained eliminating $\boldsymbol{\gamma}_{l}$ of $\boldsymbol{\Gamma}$. Note that $\boldsymbol{\gamma}_{l} \boldsymbol{\gamma}_{l}^{t} = 1$, $\boldsymbol{\Gamma}_{-l} \boldsymbol{\gamma}_{l}^{t} = \mathbf{0}$ and $\boldsymbol{\Gamma}_{-l} \boldsymbol{\Gamma}_{-l}^{t} = \mathbf{I}$, where $\mathbf{0}$ is a vector of zeros with dimension $(m-1) \times 1$ and $\mathbf{I}$ is the identity matrix with order $m-1$.

In that case, it is obtained that:
\begin{eqnarray*}
        \mathbf{x}_{a,l}^{t} \mathbf{x}_{a,l} &=& \mathbf{x}_{l}^{t} \mathbf{x}_{l} + k \boldsymbol{\gamma}_{l} \boldsymbol{\gamma}_{l}^{t} = \mathbf{x}_{l}^{t} \mathbf{x}_{l} + k, \\
        \mathbf{x}_{a,-l}^{t} \mathbf{x}_{a,l} &=& \mathbf{x}_{-l}^{t} \mathbf{x}_{l} + k \boldsymbol{\Gamma}_{-l} \boldsymbol{\gamma}_{l}^{t} = \mathbf{x}_{-l}^{t} \mathbf{x}_{l}, \\
        \mathbf{x}_{a,-l}^{t} \mathbf{x}_{a,-l} &=& \mathbf{x}_{-l}^{t} \mathbf{x}_{-l} + k \boldsymbol{\Gamma}_{-l} \boldsymbol{\Gamma}_{-l}^{t} = \mathbf{x}_{-l}^{t} \mathbf{x}_{-l} + k \mathbf{I}.
    \end{eqnarray*}

Thus, the expression (\ref{VIF_K}) can be rewritten as (for $l=1,\dots,m$):
\begin{equation}
        VIF(l, k) = \frac{\mathbf{x}_{l}^{t} \mathbf{x}_{l} + k}{\mathbf{x}_{l}^{t} \mathbf{x}_{l} + k - \mathbf{x}_{l}^{t} \mathbf{x}_{-l} \left( \mathbf{x}_{-l}^{t} \mathbf{x}_{-l} + k \mathbf{I} \right)^{-1} \mathbf{x}_{-l}^{t} \mathbf{x}_{l}}. \label{VIF_k}
    \end{equation}
This expression coincides with the one given by \cite{Rodriguezetal2021}. In that paper, the Stewart index (see \cite{Stewart1987}) is adapted to be applied in the ridge regression, and it is stated that \textit{if independent variables are centered [...], the Stewart index coincides with the VIF obtained in a centered model}.
That paper also shows the following:
\begin{itemize}
\item The expression (\ref{VIF_k}) is continuous (i.e., for $k=0$, it coincides with the value obtained by the OLS), is decreasing as a function of $k$ and always is higher than or equal to its minimum value.
\item For standardized data (the situation considered in this paper), expression (\ref{VIF_k}) coincides with the one proposed by \cite{Garciaetal2016}.
\end{itemize}

\item However, in the particular case when $\mathbf{K} = diag (0,\dots,k_{l},\dots,0)$, $l=1,\dots,m$, it is verified that:
$$\mathbf{x}_{a} = \left(
\begin{array}{c}
            \mathbf{x} \\
            \sqrt{k_{l}} \boldsymbol{\Upsilon}
        \end{array} \right), \quad
\boldsymbol{\Upsilon} = \left(
\begin{array}{ccccc}
            0 & \dots & 0 & \dots & 0 \\
            \vdots &  & \vdots &  & \vdots \\
            \gamma_{1l} & \dots & \gamma_{ll} & \dots & \gamma_{ml} \\
            \vdots &  & \vdots &  & \vdots \\
            0 & \dots & 0 & \dots & 0
        \end{array} \right)$$
$$\mathbf{x}_{a,l} = \left(
\begin{array}{c}
            \mathbf{x}_{l} \\
            \sqrt{k_{l}} \boldsymbol{\Upsilon}_{l}
        \end{array} \right), \quad
\mathbf{x}_{a,-l} = \left(
\begin{array}{c}
            \mathbf{x}_{-l} \\
            \sqrt{k_{l}} \boldsymbol{\Upsilon}_{-l}
        \end{array} \right),$$
where $\boldsymbol{\Upsilon}_{l}$ is column $l$ of $\boldsymbol{\Upsilon}$ and $\boldsymbol{\Upsilon}_{-l}$ is the result obtained when $\boldsymbol{\Upsilon}_{l}$ is eliminated from $\boldsymbol{\Upsilon}$. In that case, it is obtained that:
\begin{eqnarray*}
        \mathbf{x}_{a,l}^{t} \mathbf{x}_{a,l} &=& \mathbf{x}_{l}^{t} \mathbf{x}_{l} + k_{l} \boldsymbol{\Upsilon}_{l}^{t} \boldsymbol{\Upsilon}_{l} = \mathbf{x}_{l}^{t} \mathbf{x}_{l} + k_{l} \gamma_{ll}^{2}, \\
        \mathbf{x}_{a,-l}^{t} \mathbf{x}_{a,l} &=& \mathbf{x}_{-l}^{t} \mathbf{x}_{l} + k_{l} \boldsymbol{\Upsilon}_{-l}^{t} \boldsymbol{\Upsilon}_{l}, \\
        \mathbf{x}_{a,-l}^{t} \mathbf{x}_{a,-l} &=& \mathbf{x}_{-l}^{t} \mathbf{x}_{-l} + k_{l} \boldsymbol{\Upsilon}_{-l}^{t} \boldsymbol{\Upsilon}_{-l},
    \end{eqnarray*}
where:
$$\boldsymbol{\Upsilon}_{-l}^{t} \boldsymbol{\Upsilon}_{l} = \left(
\begin{array}{c}
            \gamma_{1l} \gamma_{ll} \\
            \vdots \\
            \gamma_{ml} \gamma_{ll}
        \end{array} \right), \quad
\boldsymbol{\Upsilon}_{-l}^{t} \boldsymbol{\Upsilon}_{-l} = \left(
\begin{array}{ccc}
            \gamma_{1l}^{2} & \dots & \gamma_{ll} \gamma_{ml} \\
            \vdots &  & \vdots \\
            \gamma_{ml} \gamma_{1l} & \dots & \gamma_{ll} \gamma_{ml}^{2}
        \end{array} \right).$$

Thus, the expression (\ref{VIF_K}) can be expressed as (for $l=1,\dots,m$):
\begin{equation}
        VIF(l, k_{l}) = \frac{\mathbf{x}_{l}^{t} \mathbf{x}_{l} + k_{l} \gamma_{ll}^{2}}{\mathbf{x}_{l}^{t} \mathbf{x}_{l} + k_{l} \gamma_{ll}^{2} - \left( \mathbf{x}_{-l}^{t} \mathbf{x}_{l} + k_{l} \boldsymbol{\Upsilon}_{-l}^{t} \boldsymbol{\Upsilon}_{l} \right)^{t} \left( \mathbf{x}_{-l}^{t} \mathbf{x}_{-l} + k_{l} \boldsymbol{\Upsilon}_{-l}^{t} \boldsymbol{\Upsilon}_{-l} \right)^{-1} \left( \mathbf{x}_{-l}^{t} \mathbf{x}_{l} + k_{l} \boldsymbol{\Upsilon}_{-l}^{t} \boldsymbol{\Upsilon}_{l} \right)}. \label{VIF_k_l}
    \end{equation}

It is evident that for $k_{l} = 0$ coincides with the expression obtained for the OLS. Furthermore, denoting $\mathbf{a} = \mathbf{x}_{-l}^{t} \mathbf{x}_{l} + k_{l} \boldsymbol{\Upsilon}_{-l}^{t} \boldsymbol{\Upsilon}_{l}$ and $\mathbf{A} = \left( \mathbf{x}_{-l}^{t} \mathbf{x}_{-l} + k_{l} \boldsymbol{\Upsilon}_{-l}^{t} \boldsymbol{\Upsilon}_{-l} \right)^{-1}$, expression (\ref{VIF_k_l}) can be rewritten as:
$$VIF(l, k_{l}) = \frac{1}{1 - \frac{\mathbf{a}^{t} \mathbf{A} \mathbf{a}}{\mathbf{x}_{l}^{t} \mathbf{x}_{l} + k_{l} \gamma_{ll}^{2}}}.$$

Since $\mathbf{A}$ is a matrix positive defined:
$$\mathbf{a}^{t} \mathbf{A} \mathbf{a} > 0 \rightarrow 1 - \frac{\mathbf{a}^{t} \mathbf{A} \mathbf{a}}{\mathbf{x}_{l}^{t} \mathbf{x}_{l} + k_{l} \gamma_{ll}^{2}} < 1 \rightarrow VIF(l, k_{l}) > 1,$$
then $VIF(l, k_{l})$ is always higher than 1 (the minimum value possible for VIF).

Finally, the analysis of the monotony is not possible due to the erratic behavior expected after the analysis of $corr( \mathbf{x}_{a,i}, \mathbf{x}_{a,j}, k_{l})$.
\end{itemize}

\section{Condition Number}\label{cn}

The condition number (CN) in the ridge regression is obtained from the eigenvalues of matrix $\mathbf{x}_{a}^{t} \mathbf{x}_{a} = \mathbf{x}^{t} \mathbf{x} + \boldsymbol{\Gamma} \mathbf{K} \boldsymbol{\Gamma}^{t}$ by using the following expression:
\begin{equation}\label{cn_K}
        CN(\mathbf{x}_{a},\mathbf{K}) = \sqrt{\frac{\mu_{max}}{\mu_{min}}},
    \end{equation}
where $\mu_{max}$ and $\mu_{min}$ are the maximum and minimum eigenvalues of $\mathbf{x}_{a}^{t} \mathbf{x}_{a}$, respectively.
\cite{Belsley1980} stated that for the calculation of the eigenvalues, the data should be transformed to unit length. This transformation is performed to ensure that if all variables are orthogonal the value of the condition number will be 1, coinciding with its minimum value possible. Other transformations, such as typification or standardization (considered in this paper), also verify this condition. \cite{Belsley1980} recommended the unit length transformation because it considers the intercept. However, in a noncentered model, this consideration is not necessary since there is no intercept. Furthermore, for a centered dataset, the unit length transformation coincides with standardization. See \cite{Salmeron2018CN} for more details about the transformation data and condition number.

Due to $\mathbf{x}^{t} \mathbf{x} = \boldsymbol{\Gamma} \boldsymbol{\Lambda} \boldsymbol{\Gamma}^{t}$, it is verified that $\mathbf{x}_{a}^{t} \mathbf{x}_{a} = \boldsymbol{\Gamma} \left( \boldsymbol{\Lambda} + \mathbf{K} \right) \boldsymbol{\Gamma}^{t}$ and, consequently, $\mu_{j} = \lambda_{j} + k_{j}$ for $j=1,\dots,m$.

Assuming that the eigenvalues are ordered as $0 < \lambda_{min} = \lambda_{(1)} < \lambda_{(2)} < \dots < \lambda_{(m)} = \lambda_{max}$, the order of the eigenvalues $0 < \mu_{min} = \mu_{(1)} < \mu_{(2)} < \dots < \mu_{(m)} = \mu_{max}$ could change depending on the values of $k_{j}$ for $j=1,\dots,m$. Thus, the behavior of $CN(\mathbf{x}_{a},\mathbf{K})$ is unclear.

\subsection{Particular cases}

\begin{figure}
\begin{center}
%\begin{minipage}[t]{6cm}
{\scriptsize
\begin{tikzpicture}

        % horizontal axis
        \draw[->] (0,0) -- (6,0) node[anchor=north] {$k_{l}$}; % \draw[->] no me funciona
        % vertical axis
        \draw[->] (0,0) -- (0,4);

        \draw (0,1) to (3,1);

%        \draw (0,2.5) .. controls (1,0.5) .. (2,0.5);
%        \draw (2,0.5) .. controls (3,2.75) .. (3.5,3.75);
        \draw (3,1) to (5.5,3.5);

        % información eje
        %\foreach \y/\ytext in {1/1, 2/2, 3/3}
        %\draw[shift={(0,\y)}] (0pt,2pt) -- (0pt,-2pt) node[below] {$\ytext$};
        \draw[fill=black]

                    (0.6,4.3) node[left] {$CN(\mathbf{x}_{a},k_{l})$}

                    (0,1) node[left] {$CN(\mathbf{x})$}

                    %(0,2.5) node[left] {$\lim_{k_{l} \rightarrow +\infty} MSE(\widehat{\boldsymbol{\beta}} (k_{l}))$}
                    %(0,2.5) node[left] {$MSE(\hat{\beta} (0))$}
                    %(0,1.75) node[left] {$MSE(\hat{\beta} (0.5+1.5h))$}
                    %(0,1) node[left] {$MSE(\widehat{\boldsymbol{\beta}} (k_{l,min}))$}
                    (3,-0.3) node[left] {$\lambda_{max} - \lambda_{l}$}

                    %(3.15,-0.3) node[left] {$0.5+1.5h$}
                    (0,0) node[left] {$0$};

                    %(0,1.25) node[right] {No tiene sentido};
        % linea horizontal discontinua
        \draw[dotted] (3,0) -- (3,1);

        %\draw[dotted] (0,2.5) -- (6,2.5);
        \draw (0,1) node {$\bullet$};

        \draw (3,0) node {$\bullet$};

        %\draw[dashed] (0,0.9) -- (1.2,0.9);
        %\draw[dashed] (1.2,0) -- (1.2,0.9);
        \draw (3,1) node {$\bullet$};

        \end{tikzpicture}}
%\end{minipage}
\caption{$CN(\mathbf{x}_{a},k_{l})$ for $\mu_{l} = \lambda_{l} + k_{l}$ and $\mu_{j} = \lambda_{j}$ with $j=1,\dots,l-1,l+1,\dots,m$}
        \label{graf_cn_2}
\end{center}
\end{figure}
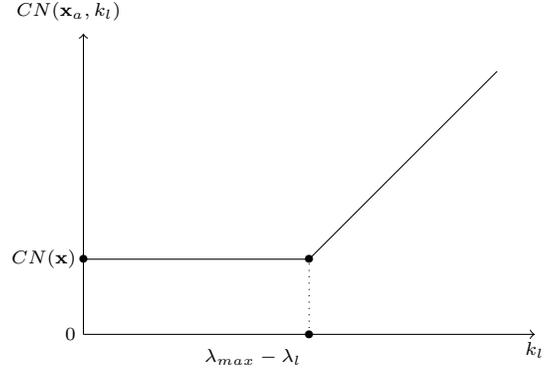

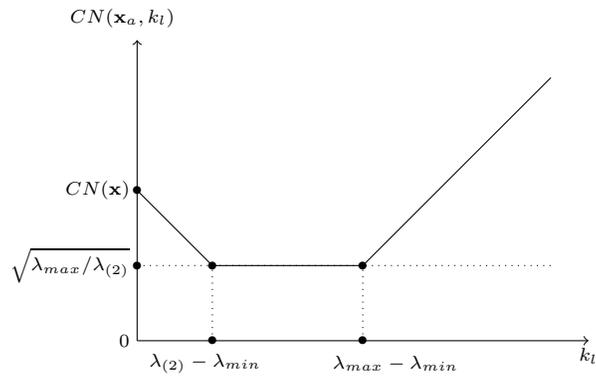
\begin{figure}
\begin{center}
%\begin{minipage}[t]{6cm}
{\scriptsize
\begin{tikzpicture}

        % horizontal axis
        \draw[->] (0,0) -- (6,0) node[anchor=north] {$k_{l}$}; % \draw[->] no me funciona
        % vertical axis
        \draw[->] (0,0) -- (0,4);

        \draw (0,2) to (1,1);

%        \draw (0,2.5) .. controls (1,0.5) .. (2,0.5);
        \draw (1,1) to (3,1);

%        \draw (2,0.5) .. controls (3,2.75) .. (3.5,3.75);
        \draw (3,1) to (5.5,3.5);

        % información eje
        %\foreach \y/\ytext in {1/1, 2/2, 3/3}
        %\draw[shift={(0,\y)}] (0pt,2pt) -- (0pt,-2pt) node[below] {$\ytext$};
        \draw[fill=black]

                    (0.6,4.3) node[left] {$CN(\mathbf{x}_{a},k_{l})$}

                    (0,2) node[left] {$CN(\mathbf{x})$}

                    %(0,2.5) node[left] {$\lim_{k_{l} \rightarrow +\infty} MSE(\widehat{\boldsymbol{\beta}} (k_{l}))$}
                    %(0,2.5) node[left] {$MSE(\hat{\beta} (0))$}
                    %(0,1.75) node[left] {$MSE(\hat{\beta} (0.5+1.5h))$}
                    (0,1) node[left] {$\sqrt{\lambda_{max}/\lambda_{(2)}}$}

                    (2.5,-0.3) node[right] {$\lambda_{max} - \lambda_{min}$}

                    (1.75,-0.3) node[left] {$\lambda_{(2)} - \lambda_{min}$}

                    (0,0) node[left] {$0$};

                    %(0,1.25) node[right] {No tiene sentido};
        % linea horizontal discontinua
        \draw[dotted] (3,0) -- (3,1);

        \draw[dotted] (0,1) -- (5.5,1);

        \draw[dotted] (1,0) -- (1,1);

        \draw (0,2) node {$\bullet$};

        \draw (1,1) node {$\bullet$};

        \draw (0,1) node {$\bullet$};

        \draw (1,0) node {$\bullet$};

        \draw (3,0) node {$\bullet$};

        %\draw[dashed] (0,0.9) -- (1.2,0.9);
        %\draw[dashed] (1.2,0) -- (1.2,0.9);
        \draw (3,1) node {$\bullet$};

        \end{tikzpicture}}
%\end{minipage}
\caption{$CN(\mathbf{x}_{a},k_{l})$ for $\mu_{min} = \lambda_{min} + k_{l}$ and $\mu_{j} = \lambda_{j}$ with $j=2,\dots,m$}
        \label{graf_cn_3}
\end{center}
\end{figure}

\begin{itemize}
\item In the particular case when $\mathbf{K} = k \mathbf{I}$, then $\mu_{min} = \lambda_{min} + k$ and $\mu_{max} = \lambda_{max} + k$. In this case, the expression (\ref{cn_K}) can be rewritten as:
\begin{equation}\label{cn_algoritmo}
        CN(\mathbf{x}_{a},k) = \sqrt{\frac{\lambda_{max} + k}{\lambda_{min} + k}}.
    \end{equation}
This expression was analyzed in Appendix B of \cite{Salmeron2018CN}, which shows that $CN(\mathbf{x}_{a},k)$ is a function continuous in $k$ (i.e., for $k=0$ coinciding with the condition number obtained for the OLS), is decreasing as a function of $k$ and is always higher than or equal to 1 (its minimum possible value).

\item  In the particular case when $\mathbf{K} = diag(0,\dots,k_{l},\dots,0)$, $l=1,\dots,m$, three scenarios are possible:
\begin{itemize}
\item If $\mu_{max} = \lambda_{max} + k_{l}$ and $\mu_{j} = \lambda_{j}$ for $j=1,\dots,m-1$, expression (\ref{cn_K}) increases as a function of $k_{l}$, i.e., the CN increases as $k_{l}$ increases. See Figure \ref{graf_cn_1}.
\item If $\mu_{l} = \lambda_{l} + k_{l}$ and  $\mu_{j} = \lambda_{j}$ for $j=1,\dots,l-1,l+1,\dots,m$, the CN is constant and equal to the one obtained for the OLS until $\mu_{l} > \mu_{max}$ (it occurs when $k_{l} > \mu_{max} - \lambda_{l}$). From this point, the CN increases as $k_{l}$ increases. See Figure \ref{graf_cn_2}.
\item If $\mu_{min} = \lambda_{min} + k_{l}$ and $\mu_{j} = \lambda_{j}$ for $j=2,\dots,m$, expression (\ref{cn_K}) is decreasing as a function of $k_{l}$ until $\mu_{min} > \lambda_{(2)}$ (it occurs when $k_{l} > \lambda_{(2)} - \lambda_{min}$). From this point, it will be constant and equal to $\sqrt{\lambda_{max}/\lambda_{(2)}}$. Finally, when $\mu_{min} = \lambda_{min} + k_{l}$ is higher than $\mu_{max} = \lambda_{max}$ (it occurs when $k_{l} > \lambda_{max} - \lambda_{min}$), expression (\ref{cn_K}) is increasing as a function of $k_{l}$. See Figure \ref{graf_cn_3}.
\end{itemize}
In all cases, $CN(\mathbf{x}_{a},k_{l})$ is continuous in $k_{l}$. I.e., for $k_{l} = 0$ coinciding with the condition number obtained in the OLS, $CN(\mathbf{x})$, being always higher than 1 (its minimum value). Thus, analogous to the behavior described for the coefficient of correlation, for high values of $k_{l}$, the degree of multicollinearity increases as $k_{l}$ increases.
\end{itemize}
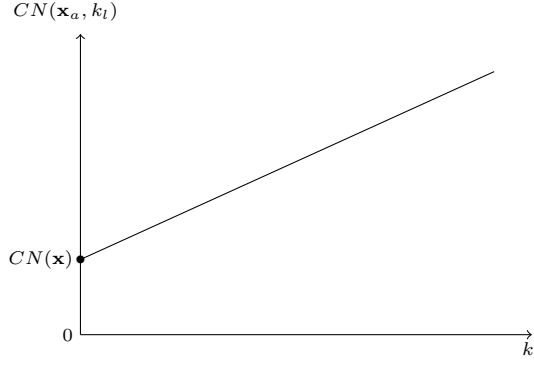
\begin{figure}
\begin{center}
%\begin{minipage}[t]{6cm}
{\scriptsize
\begin{tikzpicture}

        % horizontal axis
        \draw[->] (0,0) -- (6,0) node[anchor=north] {$k_{l}$}; % \draw[->] no me funciona
        % vertical axis
        \draw[->] (0,0) -- (0,4);

        \draw (0,1) to (5.5,3.5);

%        \draw (0,2.5) .. controls (1,0.5) .. (2,0.5);
%        \draw (2,0.5) .. controls (3,2.75) .. (3.5,3.75);
        %\draw (2.5,1.75) to[out=46,in=180] (5.5,2.43);
        % información eje
        %\foreach \y/\ytext in {1/1, 2/2, 3/3}
        %\draw[shift={(0,\y)}] (0pt,2pt) -- (0pt,-2pt) node[below] {$\ytext$};
        \draw[fill=black]

                    (0.6,4.3) node[left] {$CN(\mathbf{x}_{a},k_{l})$}

                    (0,1) node[left] {$CN(\mathbf{x})$}

                    %(0,2.5) node[left] {$\lim_{k_{l} \rightarrow +\infty} MSE(\widehat{\boldsymbol{\beta}} (k_{l}))$}
                    %(0,2.5) node[left] {$MSE(\hat{\beta} (0))$}
                    %(0,1.75) node[left] {$MSE(\hat{\beta} (0.5+1.5h))$}
                    %(0,1) node[left] {$MSE(\widehat{\boldsymbol{\beta}} (k_{l,min}))$}
                    %(1.75,-0.3) node[left] {$k_{l,min}$}
                    %(3.15,-0.3) node[left] {$0.5+1.5h$}
                    (0,0) node[left] {$0$};

                    %(0,1.25) node[right] {No tiene sentido};
        % linea horizontal discontinua
        %\draw[dotted] (0,3) -- (6,3);
        %\draw[dotted] (0,2.5) -- (6,2.5);
        \draw (0,1) node {$\bullet$};

        %\draw[dashed] (0,0.9) -- (1.2,0.9);
        %\draw[dashed] (1.2,0) -- (1.2,0.9);
        %\draw (1.2,0.9) node {$\bullet$};
        \end{tikzpicture}}
%\end{minipage}
\caption{$CN(\mathbf{x}_{a},k_{l})$ for $\mu_{max} = \lambda_{max} + k_{l}$ and $\mu_{j} = \lambda_{j}$ with $j=1,\dots,m-1$}
        \label{graf_cn_1}
\end{center}
\end{figure}

Finally, from the third scenario, it is deduced that the value of $l$ should correspond to the minimum eigenvalue since for $\lambda_{(2)} - \lambda_{min} < k_{l} < \lambda_{max} - \lambda_{min}$, the minimum value $CN(\mathbf{x}_{a},k_{l})$ that coincides with $\sqrt{\lambda_{max}/\lambda_{(2)}}$ is obtained.

Thus, the value of $l$ is known once the eigenvalues and the values of $k_{l}$ that obtain the minimum degree of multicollinearity are known.

\section{Examples}
    \label{example}

To illustrate the results obtained in this paper, two real data sets will be analysed.

The first example illustrates the possibilities for mitigating multicollinearity in a multiple linear regression model by means of the results obtained in Section \ref{cn}.

On the other hand, \cite{SalmeronGRR2021} compares the results obtained with those provided by the R packages \textit{lmridge} (\cite{lmridge,lmridgeR}) and \textit{lrmest} (\cite{lrmest}), obtaining that their results coincide with those provided by the latter for the regular case and differ from those obtained from the former.
For this reason, in the second example, the comparison of the proposals made in this work (and in \cite{SalmeronGRR2021}) and those obtained from the \textit{lmridge} package are discussed in more detail. In addition, in order to enrich the study, the \textit{ridge} (\cite{ridge}) package is added to the comparison.

The code utilized in R (\cite{RCoreTeam}) to generate the results presented in both examples is accessible on GitHub, specifically at \url{https://github.com/rnoremlas/GRR/tree/main/02_Nonorthogonal_models}.

\subsection{Example 1}

To illustrate the contribution of this paper, we use the dataset previously applied by \cite{Salmeron2020} with the following information about 15 Spanish companies: number of employees, $\mathbf{E}$ (dependent variable), fixed assets, $\mathbf{FA}$, operating income, $\mathbf{OI}$, and sales, $\mathbf{S}$ (all variables measured in euros).

The matrix of the correlation and its determinants are presented as follows:

$$\mathbf{R} = \left(
\begin{array}{ccc}

            1 & 0.7264656 & 0.7225473 \\

            0.7264656 & 1 & 0.9998871 \\

            0.7225473 & 0.9998871 & 1

        \end{array} \right), \quad det(\mathbf{R}) = 0.00009190317.$$

Table \ref{tabla1} shows the coefficients of variation and variance inflation factors and Table \ref{tabla2} shows the condition number with and without considering the intercept. All the results are obtained for the original data (subindex $o$) and standardized data (subindex $s$), except for the matrix of correlation, its determinants and the variance inflation factors because they are invariant to origin and scale transformations (see, for example, \cite{Garciaetal2016}).

\begin{table}

        \centering

        \begin{tabular}{cccc}

            \hline

            Measure & Fixed Assets & Operating Income & Sales \\

            \hline

            $CV_{o}$ & 0.9806092 & 0.7216219 & 0.7199330 \\

            $CV_{s}$   & 2.567155$\cdot 10^{16}$ & 5.441274$\cdot 10^{16}$ & 1.975226$\cdot 10^{16}$ \\

            $VIF$ & 2.45664 & 5200.31530 & 5138.53548 \\

            \hline

        \end{tabular}

        \caption{Coefficient of Variation and Variance Inflation Factors for Fixed Assets, Operating Income and Sales} \label{tabla1}

    \end{table}

\begin{table}

        \centering

        \begin{tabular}{cccc}

            \hline

            Measure & Without Intercept & With Intercept \\

            \hline

            $CN_{o}$ & 289.9235 & 327.6666 \\

            $CN_{s}$   & 165.2056 & 165.2056 \\

            \hline

        \end{tabular}

        \caption{ condition number for Fixed Assets, Operating Income and Sales} \label{tabla2}

    \end{table}

From these measures (see, for example, \cite{multiColl1} and \cite{multiColl2}), it is possible to conclude the following:
\begin{itemize}
\item The matrix of correlation, its determinants and the variance inflation factors indicate that the model presents a troubling degree of essential multicollinearity with a strong linear relationship between variables $\textbf{OI}$ and $\textbf{S}$.
\item The condition number for the original data also indicates that the multicollinearity is troubling. To determine whether multicollinearity is essential or nonessential, we could use the decomposition of the variance proposed by \cite{Belsley1980}. However, there is a small difference between the CNs considering and not considering the intercept, which suggests that the role of the intercept is irrelevant and, consequently, that multicollinearity will be essential.
\item The CN for standardized data indicates that the degree of essential multicollinearity is troubling.
\item The coefficients of variation confirm that the nonessential multicollinearity of the original data is not troubling. With standardized data, the coefficients of variation should be infinite (in any case, they are very high).
\end{itemize}
Thus, when original data are used, the conclusion of the diagnosis is that the model presents essential multicollinearity. This problem is not mitigated with the standardization of the data. For this reason, we apply the regular and generalized versions of the ridge regression.

Assuming that data are standardized, the eigenvalues of matrix $\mathbf{x}^{t} \mathbf{x}$, where $\mathbf{X} = [ \textbf{FA}, \textbf{OI}, \textbf{S} ]$, are:
$$2.640016, 0.3598875, 0.00009672911.$$
Considering the results shown in Section \ref{cn}, it is concluded that the third variable is the variable that should be transformed in the generalized ridge regression (i.e., $l=3$). In this case, the minimum CN is equal to 2.708444, which is obtained when $0.3597908 < k_{3} < 2.639919$.

The same election of $l$ is obtained (see Table \ref{tabla3}) if we want to minimize the mean squared error (MSE) by following \cite{SalmeronGRR2021} since the minimum MSE, 190.9562, is obtained for $k_{3,min} = 0.00003485569$ (note that the MSE of the OLS is equal to 259.7374). Because the condition number is high for $k_{3,min}$ in the regular (141.6457) and generalized (141.6447) cases, it is interesting to analyze whether there is a value of $k$ or $k_{3}$ that obtains a CN lower than the recommended threshold and, at the same time, provides an MSE lower than the one provided by the OLS.

\begin{table}
        \centering
        \begin{tabular}{ccc} % c}
            \hline
            l & $k_{l,min}$ & MSE for $k_{l,min}$ \\ % & ¿Ridge MSE is always lower than OLS MSE? \\
            \hline
            1 & 0.3144335 & 259.7364 \\ %  & FALSE \\
            2 & 0.02162365 & 259.7335 \\ %  & FALSE \\
            3 & 0.00003485569 & 190.9562 \\ %  & FALSE \\
            \hline
        \end{tabular}
        \caption{$k_{l,min}$ election} \label{tabla3}
\end{table}

\begin{figure}
          \centering
          \includegraphics[width=9cm]{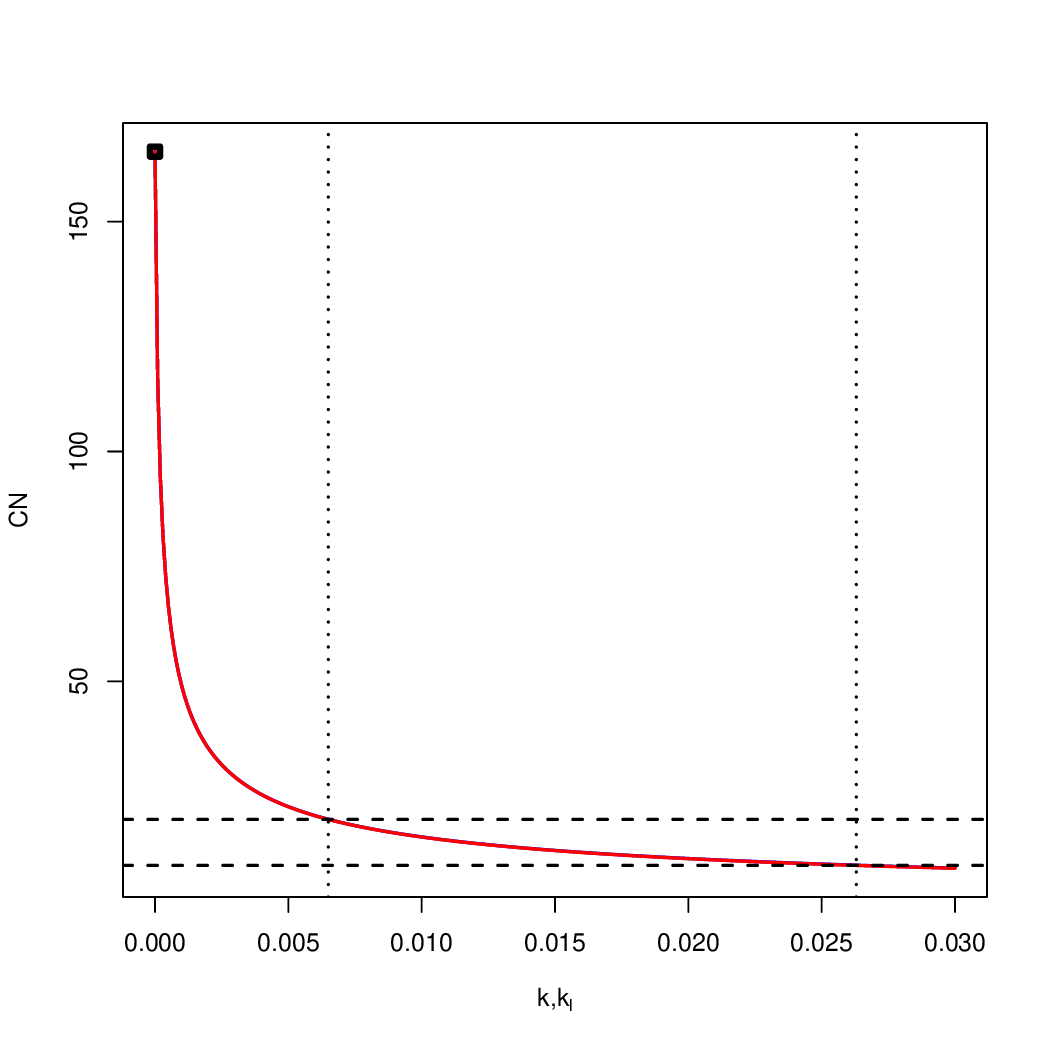}
          \includegraphics[width=9cm]{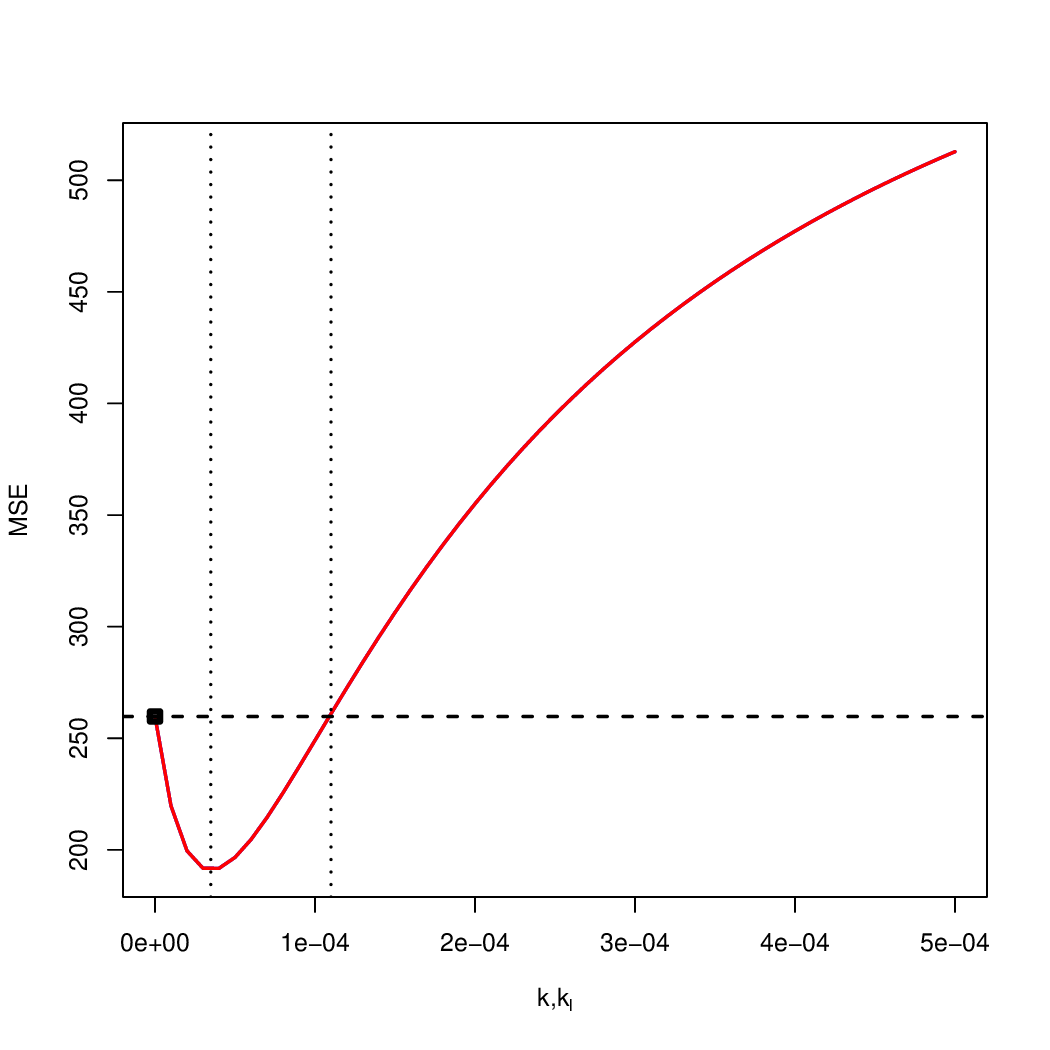}\\
          \caption{Condition number and mean squared error for $k, k_{3} \in [0, 0.03]$ and $k, k_{3} \in [0, 0.0005]$, respectively} \label{Figura1}
\end{figure}

Figure \ref{Figura1} shows the CN and the MSE for the regular and generalized ridge regressions. Note that the results for both are practically the same. Looking at the top of the figure, note that the CN is lower than 20 when $k > 0.00651$ or $k_{3} > 0.0065$ and lower than 10 if $k > 0.02656$ or $k_{3} > 0.0263$. Looking at the bottom of the figure, note that the MSE of the ridge regression is lower than that obtained from the OLS when $k, k_{3} < 0.00011$. These values were calculated using Algorithms \ref{algoritmo1} and \ref{algoritmo2}, respectively.

\begin{algorithm}
\begin{algorithmic}[1]
\REQUIRE Calculate $\lambda_{1},\dots,\lambda_{m}$, $D(n)$ := \{discretization of the interval [0,1] with $n$ points \} and determinate the value \textit{threshold} (10 or 20)
     \STATE j = 1
\FOR {$k, k_{l} \in D(n)$}
        \STATE Calculate $CN \left( \mathbf{x}_{a}, \mathbf{K} \right)$ with expression (\ref{cn_algoritmo}) and subsequent indications and store it in $cn_{j}$
%        \IF {$j > 1$}
            \IF {$cn_{j} < threshold$}
                \STATE index = j
                \STATE break
            \ENDIF
%        \ENDIF
        \STATE j = j + 1
\ENDFOR
     \STATE $k, k_{l} = D[index]$
\end{algorithmic}
\caption{Calculation of $k$ and $k_{l}$ that coincide with a given $CN \left( \mathbf{x}_{a}, \mathbf{K} \right)$} \label{algoritmo1}
\end{algorithm}

\begin{algorithm}
\begin{algorithmic}[1]
\REQUIRE Calculate $\mathbf{K}$, $\boldsymbol{\Omega}$, $\boldsymbol{\Psi}$, $\boldsymbol{\Theta}$, MSE in OLS (\textit{mse.ols}) and $D(n)$ := \{discretization of the interval [0,1] with $n$ points \}
     \STATE j = 1
\FOR {$k, k_{l} \in D(n)$}
        \STATE Calculate $MSE \left( \widehat{\boldsymbol{\beta}}(\mathbf{K}) \right)$ with the expression given in Section \ref{ndt} and store it in $mse_{j}$
        \IF {$j > 1$}
            \IF {$mse_{j} > mse.ols$}
                \STATE index = j-1
                \STATE break
            \ENDIF
        \ENDIF
        \STATE j = j + 1
\ENDFOR
     \STATE $k, k_{l} = D[index]$
\end{algorithmic}
\caption{Calculation of $k$ and $k_{l}$ such that $MSE \left( \widehat{\boldsymbol{\beta}}(\mathbf{K}) \right) < MSE \left( \widehat{\boldsymbol{\beta}} \right)$} \label{algoritmo2}
\end{algorithm}

Thus, it is infeasible to find values of $k$ and $k_{3}$ that mitigate the multicollinearity and, at the same time, provide a MSE lower than that obtained by OLS.
However, considering that CN decreases until the value 0.3597908 and the MSE is higher than the one obtained by the OLS starting at 0.00011, it is possible to consider $k=0.00652$ and $k_{3}=0.00651$ as appropriate values since CN is lower than 20 (the threshold traditionally recommended) in these cases. These are the lowest values of $k$ and $k_{3}$ from which the multicollinearity is considered not troubling with the lowest MSE.

This method used to select $k$ and $k_{3}$ is now compared with other methods applied in the scientific literature. Thus, we obtain the estimated coefficients, the mean squared error and the goodness of fit proposed by \cite{SalmeronGRR2021} for the following cases:
\begin{enumerate}[a)]
\item (OLS) $\mathbf{K} = diag(0,\dots,0)$; and
\item (RR) $\mathbf{K} = k \mathbf{I}$ where $k = p \cdot \frac{\sigma^{2}}{\boldsymbol{\beta}^{t} \boldsymbol{\beta}} = k_{HKB}$ (\cite{HoerlKennardBaldwin}), $k = \frac{\sigma^{2}}{\xi_{max}^{2}} = k_{HK}$ (\cite{HoerlKennard1970b}) and $k = k_{min}$ that truly minimize $MSE \left( \widehat{\boldsymbol{\beta}}(k) \right)$, obtained using Algorithm 1 detailed in \cite{SalmeronGRR2021}.
\end{enumerate}
For all the calculations, we use the estimation of $\sigma^{2}$ (0.0251165) and $\boldsymbol{\beta}$ (-0.6076545, -18.4692986 and 19.5023718) obtained by the OLS.
The results are summarized in Table \ref{tabla4}.

\begin{sidewaystable}
        \centering
\begin{tabular}{ccccccc}
            \hline
$\mathbf{K}$ & OLS & $k_{HKB} = 0.0001043872$ & $k_{min} = 0.00004$ & $k_{HB} = k_{3,min} = 0.00003485569$ & $k = 0.00652$ & $k_{3} = 0.00651$ \\
            \hline
\multirow{2}{*}{$\textbf{FA}$} &  -0.6076545  &  -0.6886758  &    -0.6533678   &     -0.649136   &   -0.7460679   &    -0.7619596  \\
&   (0.2483993) &    (0.2342655) &       (0.239326) &       (0.2400644)  &    (0.2260173)   &      (0.2299292)  \\
\multirow{2}{*}{$\textbf{OI}$} & -18.4692986  &  -8.5881095 &   -12.8998793  &    -13.426373  &    0.2825805  &     0.2896735   \\
&  (11.4286357) & (5.4976346) &     (8.085598) &     (8.4016426)   &    (0.2013429)  &     (0.2024289)   \\
\multirow{2}{*}{$\textbf{S}$} &   19.5023718  &   9.6798295   &   13.9660526     &    14.489503    &   0.8470968   &     0.8552064  \\
&    (11.3605467) &     (5.4649252) &      (8.037445)  &      (8.3516049)  &    (0.2016539)  &     (0.2027776)  \\
            \hline
$MSE$ & 259.7374 & 254.2711 & 191.706 & 190.9562 & 699.8037 & 699.7746 \\
$GoF$ & 0.698602  & 0.6798243  & 0.6926366  & 0.6937112 & 0.6307898  & 0.6309265  \\
            \hline
$CN$ &  165.2056  &  114.5746  & 138.9556   &  141.6457  & 19.99941   &  19.98987  \\
            \hline
\end{tabular}
\caption{Calculation of the generalized ridge estimators, standard deviations, mean squared errors and goodness of fits for different values of $\mathbf{K}$} \label{tabla4}
\end{sidewaystable}

Note that the lowest MSE corresponds to $k_{3,min}$, which coincides with $k_{HB}$.
Furthermore, this value also provides the highest goodness of fit after the OLS. However, the condition number presents a high value. For values that lead to a CN lower than 20, the MSE is the highest, although the standard deviations are the lowest. In both cases, there is a relevant change in the estimated coefficients of variables $\textbf{OI}$ and $\textbf{S}$ (which were the ones that present the strongest linear relationship): the unexpected negative sign of $\textbf{OI}$ is corrected.

\begin{figure}
          \centering
          \includegraphics[width=9cm]{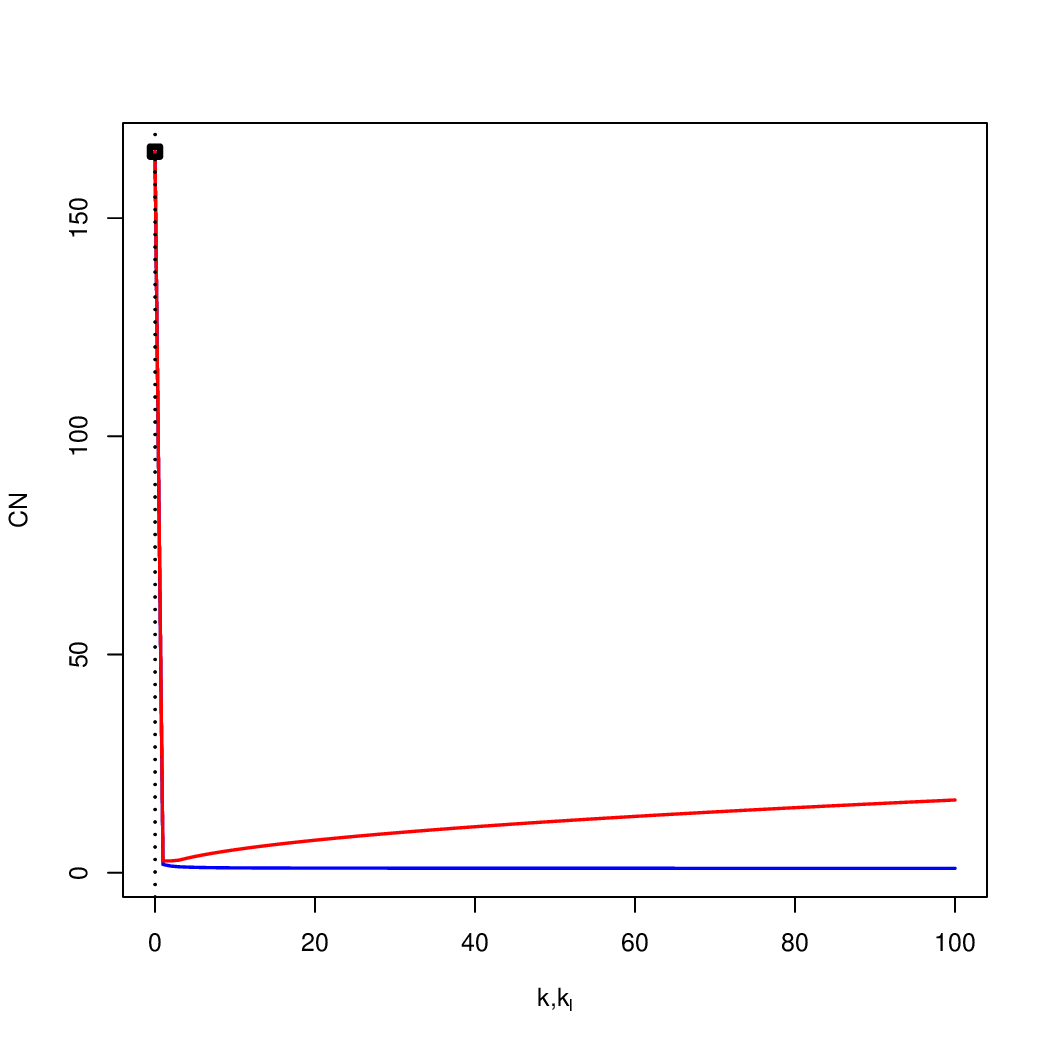}\\
\caption{Conditions number for $k, k_{3} \in [0, 100]$} \label{Figura2}
\end{figure}

Next, we analyze the asymptotic behavior of these measures. Figure \ref{Figura2} shows the calculation of CN for the regular (blue) and generalized (red) ridge regressions. As previously commented, their values are similar for the initial values, but posteriorly, the generalized CN shows a tendency to increase as $k_{3}$ increases.

\begin{figure}
          \centering
          \includegraphics[width=9cm]{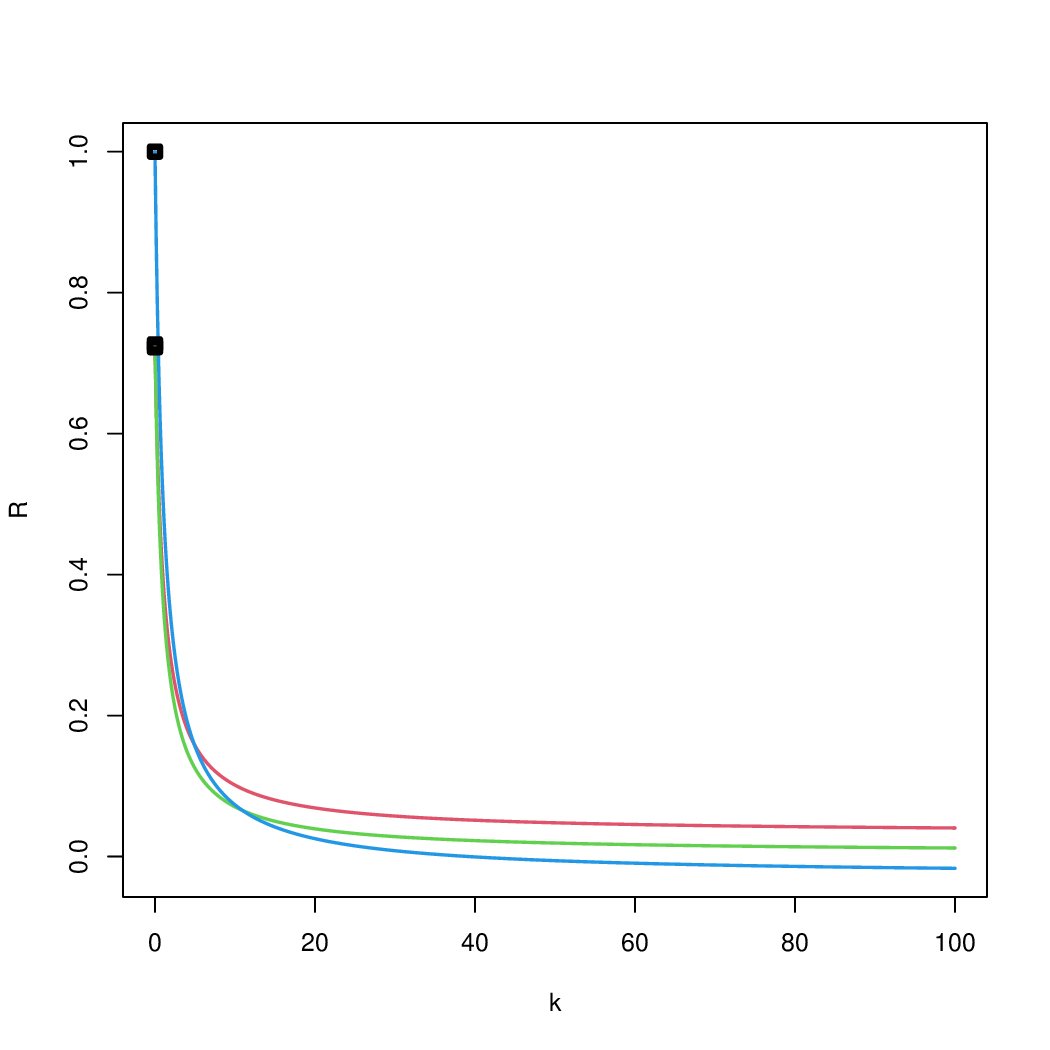}
          \includegraphics[width=9cm]{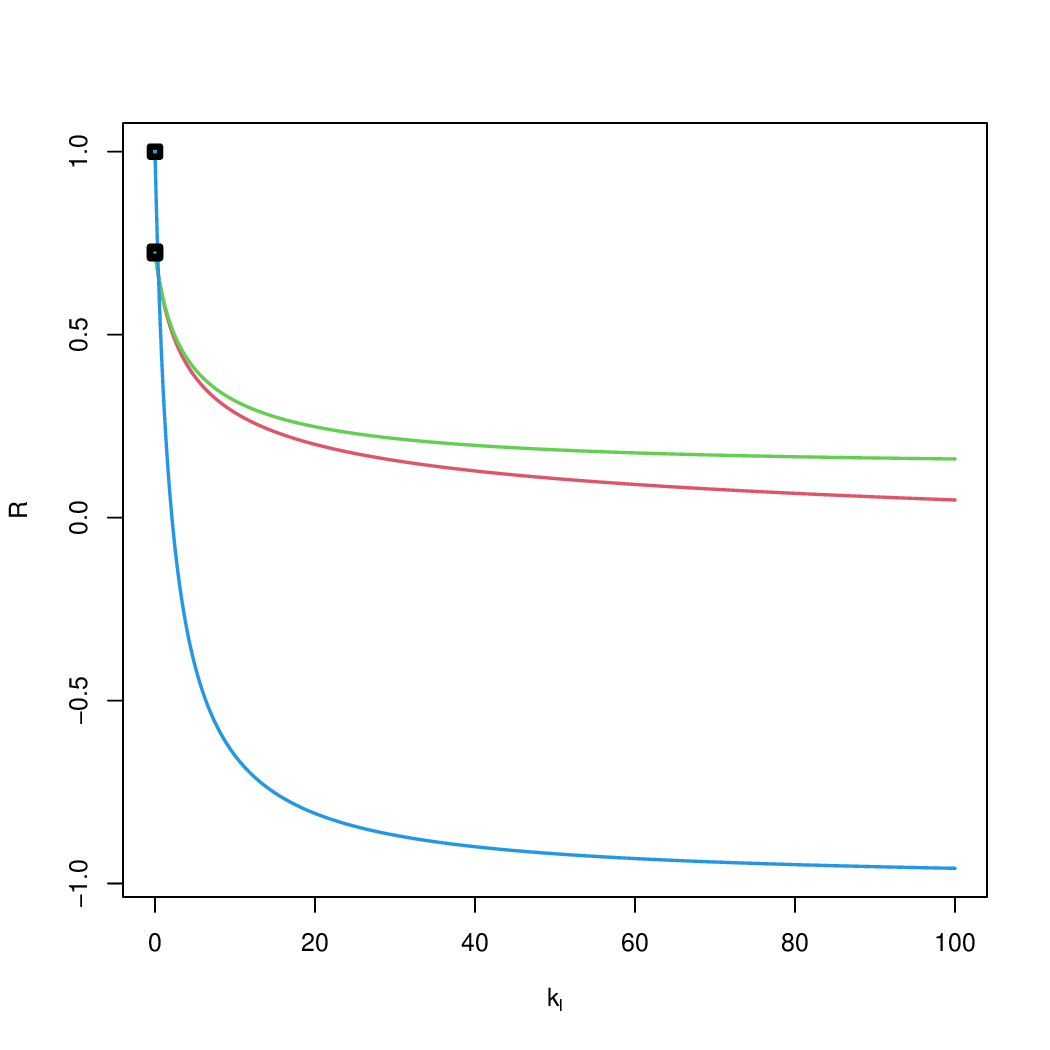}\\
\caption{Coefficients of correlation for the regular (top) and generalized (bottom) cases for $k, k_{3} \in [0, 100]$} \label{Figura3}
\end{figure}

\begin{figure}
          \centering
          \includegraphics[width=9cm]{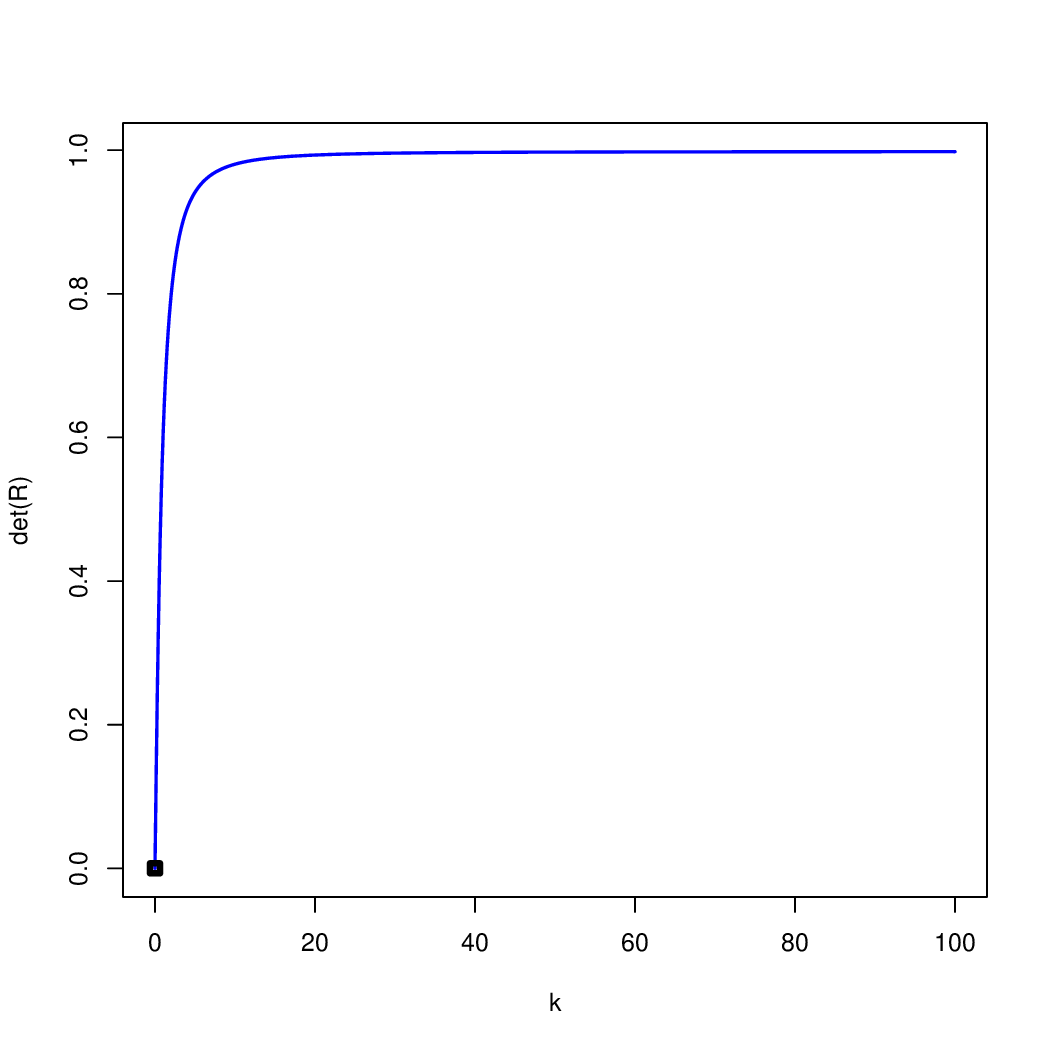}
          \includegraphics[width=9cm]{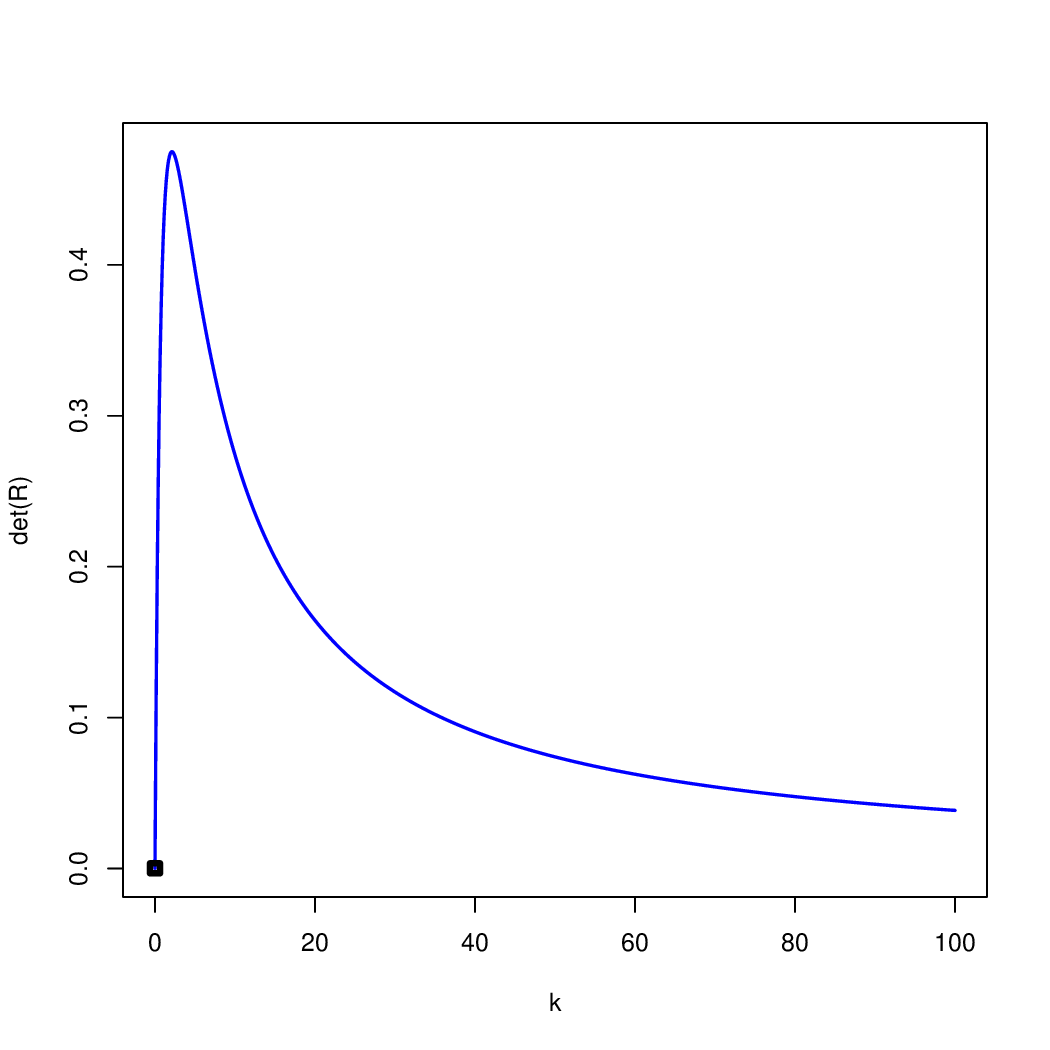}\\
\caption{Determinants of the matrix of correlations for the regular (top) and generalized (bottom) cases for $k, k_{3} \in [0, 100]$} \label{Figura4}
\end{figure}

Figure \ref{Figura3} shows the coefficients of correlation for the regular (top) and generalized (bottom) ridge regressions. In the regular case, the coefficients of correlation tend to zero; while in the generalized case, one of the coefficients of correlation tends to -1, suggesting a tendency to approach perfect multicollinearity. Furthermore, Figure \ref{Figura4} shows the determinant of the matrix of correlation. For the regular case (top), the determinant tends to one (which is associated with no correlations); while in the generalized case, the determinant presents a tendency to decrease for high values of $k_{3}$ (suggesting an increase in the degree of multicollinearity).

Finally, Tables \ref{tabla5} and \ref{tabla6} show the values of the VIFs for the regular and generalized cases, respectively. Analogous to what happens with CN, note that the behavior is similar for the initial values. However, for high values of $k$ and $k_{3}$, the VIFs increase in the generalized case but tend to one in the regular case.
\begin{table}

        \centering

        \begin{tabular}{cccc}

            \hline

            $k$ & $VIF(1,k)$ & $VIF(2,k)$ & $VIF(3,k)$ \\

            \hline

            0 & 2.456640  & 5200.315301 &  5138.535476 \\

            0.01 & 2.074291  &   50.824611  &   50.245183 \\

            0.02 & 2.040503   &  26.033831  &   25.748656 \\

            0.03 & 2.009505  &   17.713846  &   17.527190 \\

            0.04 & 1.980338  &   13.541571  &   13.404158 \\

            0.05 & 1.952719  &   11.033336  &   10.925407 \\

            0.06 & 1.926490  &    9.358576  &    9.270241 \\

            0.07 & 1.901533   &   8.160668  &    8.086274 \\

            0.08 & 1.877751   &   7.261080   &   7.197095 \\

            0.09 & 1.855061   &   6.560534  &    6.504607 \\

            0.1 & 1.833387   &   5.999423   &   5.949907     \\

            10 & 1.008017   &   1.012030   &   1.011982 \\

            20 & 1.002278   &   1.003366   &   1.003353 \\

            30 & 1.001059   &   1.001558   &   1.001552 \\

            40 & 1.000610   &   1.000895   &   1.000891 \\

            50 & 1.000396   &   1.000580   &   1.000578 \\

            60 & 1.000278   &   1.000406   &   1.000405 \\

            70 & 1.000205   &   1.000300   &   1.000299 \\

            80 & 1.000158    &  1.000231   &   1.000230 \\

            90 & 1.000125    &  1.000183   &   1.000182 \\

            100 & 1.000102    &  1.000149   &   1.000148   \\

            \hline

        \end{tabular}

        \caption{VIFs for the regular ridge regression} \label{tabla5}

    \end{table}

\begin{table}

        \centering

        \begin{tabular}{cccc}

            \hline

            $k_{3}$ & $VIF(1,k_{3})$ & $VIF(2,k_{3})$ & $VIF(3,k_{3})$ \\

            \hline

            0 & 2.456640  & 5200.315301 &  5138.535476 \\

            0.01 & 2.108184  &   50.585355  &   50.006325 \\

            0.02 & 2.106508  &   25.801383  &   25.516986 \\

            0.03 & 2.105946  &   17.488684  &   17.303170 \\

            0.04 & 2.105665  &   13.323604  &   13.187683 \\

            0.05 & 2.105496  &   10.822377  &   10.716275 \\

            0.06 & 2.105384   &   9.154416  &    9.068230 \\

            0.07 & 2.105304   &   7.963099   &   7.891165 \\

            0.08 & 2.105244   &   7.069901   &   7.008676 \\

            0.09 & 2.105197   &   6.375555   &   6.322676 \\

            0.1 & 2.105160   &   5.820464   &   5.774275 \\

            10 & 2.105533  &    3.419870  &    3.495073 \\

            20 & 2.106247   &   5.994446  &    6.132851 \\

            30 & 2.106963   &   8.585787  &    8.787195 \\

            40 & 2.107679   &  11.181319  &   11.445681 \\

            50 & 2.108395  &   13.778527  &   14.105822 \\

            60 & 2.109111  &   16.376573  &   16.766793 \\

            70 & 2.109828  &   18.975099  &   19.428236 \\

            80 & 2.110544  &   21.573924  &   22.089975 \\

            90 & 2.111260   &  24.172948  &   24.751912 \\

            100 & 2.111976  &   26.772112  &   27.413986 \\

            \hline

        \end{tabular}

        \caption{VIFs for the generalized ridge regression} \label{tabla6}

    \end{table}

\subsection{Example 2}

In this example we will work with data from \cite{Longley1967} which has the following six economic variables observed annually from 1947 to 1962 (16 observations): GNP implicit price deflator (\textbf{GNP deflator}, 1954=100), Gross National Product (\textbf{GNP}), number of unemployed (\textbf{Unemployed}), number of people in the armed forces (\textbf{Armed Forces}), non-institutionalized population greater than or equal to 14 years of age  (\textbf{Population}) and number of people employed (\textbf{Employed}, dependent variable).

Table \ref{tabla_example2_OLS} shows the results of estimating the model by OLS with standardized data. It can be observed that only the coefficient associated with the variable \textbf{GNP} is significantly different from zero and that the model is jointly valid (all at 5\% significance).

    \begin{table}
        \centering
        \begin{tabular}{ccc}
            \hline
            Variable & Estimation & p-value \\
            \hline
            GNP deflator & -0.14892 & 0.7081   \\
            GNP  & 2.03784    & 0.0365  \\
            Unemployed & -0.10746 & 0.3548 \\
            Armed Forces & -0.11107 & 0.0626  \\
            Population & -0.79922 & 0.2264       \\
            \hline
            R-squared &  \multicolumn{2}{c}{0.9874} \\
            F-statistic (p-value) & \multicolumn{2}{c}{172.1 (4.665 $\cdot 10^{-10}$)} \\
            \hline
        \end{tabular}
        \caption{OLS estimation for Longley data \cite{Longley1967} } \label{tabla_example2_OLS}
    \end{table}

The condition number obtained is equal to 61.5302. Therefore, it can be established that the degree of approximate multicollinearity in the model is troubling.

To mitigate this problem, we start by considering the following situations:
\begin{enumerate}[a)]
\item (OLS) $\mathbf{K} = diag(0,\dots,0)$;
\item (RR) $\mathbf{K} = k \mathbf{I}$ where $k = p \cdot \frac{\sigma^{2}}{\boldsymbol{\beta}^{t} \boldsymbol{\beta}} = k_{HKB}$ and $k = \frac{\sigma^{2}}{\xi_{max}^{2}} = k_{HK}$;
\item (GR) $\mathbf{K} = diag(0,\dots,k_{l,min},\dots,0)$ where $k_{l,min}$ is determined as was specified in \cite{SalmeronGRR2021}; and
\item (GR) $\mathbf{K} = diag(0,\dots,k_{l},\dots,0)$ where $k_{l}$ is determined as was specified in this paper.
\end{enumerate}
For all the calculations, we use the estimation of $\sigma^{2}$ (0.001147482) and $\boldsymbol{\beta}$ (-0.1489163,  2.0378367, -0.1074620, -0.1110659 and -0.7992234) obtained by the OLS estimation. Then it follows that $k_{HKB} = 0.0011860044$ and $k_{HK} = 0.0002705469$.

\begin{table}
        \centering
        \begin{tabular}{ccc} % c}
            \hline
            l & $k_{l,min}$ & MSE for $k_{l,min}$ \\ % & ¿Ridge MSE is always lower than OLS MSE? \\
            \hline
            1 & 0.0045807828  & 1.287685 \\ %  & FALSE
            2 & 0.0918118550  & 1.287615 \\ %  & FALSE
            3 & 0.0035558367  & 1.287584 \\ %  & FALSE
            4 & 0.1087330648  & 1.219864 \\ %  &  TRUE
            5 & 0.0002705469  & 1.021660 \\ %  & FALSE
            \hline
        \end{tabular}
        \caption{$k_{l,min}$ election} \label{tabla_example2_klmin}
\end{table}

Following \cite{SalmeronGRR2021}, see Table \ref{tabla_example2_klmin}, the lowest MSE is obtained from $k_{5,min} = 0.0002705469$.
Taking into account that the eigenvalues of the independent variables correspond to:
$$3.6096690885, 1.1753398694, 0.1991553621, 0.0148822458, 0.0009534342,$$
it is obtained $l=5$, so that the minimum value for the condition number, 15.57397, is obtained when $0.01392881 \leq k_{5} \leq 3.60871565$. For this reason, in case d) mentioned above, it would be considered that $k_{5} = 0.01392881$ and $k_{5} = 3.60871565$.

The results obtained are summarized in Table \ref{tabla_example2_results}. In this case, confidence regions are provided based on the bootstraps procedure specified in \cite{SalmeronGRR2021}.

\begin{sidewaystable}
        \centering
        \begin{tabular}{ccccccc}
             & $k = 0$ & $k_{HKB} = 0.0011860044$ & $k_{HK} = 0.0002705469$ & $k_{5,min} = 0.0002705469$ & $k_{5} = 0.01392881$ & $k_{5} = 3.60871565$ \\
            \hline
            \multirow{2}{*}{GNP deflator} & -0.1489 & 0.1789 & -0.0172 &  -0.0156 & 0.4157 & 0.4542 \\
                & (-1.0386, 0.9278) & (-0.3016, 0.7301) & (-0.5988, 0.7884) &  (-0.6233, 0.7942) & (-0.2684, 0.8463) & (-0.246, 0.8734) \\
            \multirow{2}{*}{GNP} & 2.0378 & 1.1478 & 1.683 &  1.683 & 0.5354 & 0.433 \\
                & (-0.3464, 4.1346) & (0.2928, 1.7034) & (0.0478, 2.8344) & (0.0462, 2.8374) & (0.3984, 0.8235) & (0.3603, 0.736) \\
            \multirow{2}{*}{Unemployed} & -0.1075 & -0.2043 & -0.1465 & -0.147 & -0.2749 & -0.2863 \\
                & (-0.4182, 0.139) & (-0.3853, -0.118) & (-0.3874, -0.0096) & (-0.3903, -0.0101) & (-0.4504, -0.1825) & (-0.4614, -0.19) \\
            \multirow{2}{*}{Armed Forces} & -0.1111 & -0.112 & -0.1118 & -0.1124 & -0.1166 & -0.117 \\
                & (-0.2801, -0.0483) & (-0.2593, -0.0517) & (-0.2698, -0.0541) & (-0.2735, -0.0541) & (-0.2797, -0.0417) & (-0.2802, -0.0396) \\
            \multirow{2}{*}{Population} & -0.7992 & -0.1711 & -0.5494 & -0.5503 & 0.2549 & 0.3267 \\
                & (-2.3425, 0.8616) & (-0.6806, 0.5708) & (-1.5001, 0.6404) & (-1.503, 0.6474) & (-0.0679, 0.8087) & (-0.0167, 0.8669) \\
            \hline
            MSE & 1.2877 & 1.6156 & 1.0189 & 1.0217 & 3.8044 & 4.3233 \\
            \multirow{2}{*}{GoF} & 0.9874 & 0.9861 & 0.9872 & 0.9872 & 0.9838 & 0.9833 \\
                & (0.9787, 0.9985) & (0.9749, 0.9979) & (0.9781, 0.9983) & (0.9781, 0.9983) & (0.9686, 0.9977) & (0.9673, 0.9977) \\
            CN & 61.5302 & 41.0823 & 54.3078 & 54.3058 & 15.574 & 15.574 \\
            \hline
        \end{tabular}
        \caption{Calculation of the generalized ridge estimators, bootstraps inference, mean squared errors, goodness of fits and condition number for different values of $\mathbf{K}$} \label{tabla_example2_results}
\end{sidewaystable}

It can be observed that, contrary to what is indicated in Table\ref{tabla_example2_OLS}, for $k=0$ the only coefficient significantly different from zero is that corresponding to the variable \textbf{Armed Forces}.

In this case, the lowest MSE corresponds to $k = k_{HK}$, 1.0189, very close to that obtained by $k = k_{5, min}$, 1.0217. In both cases, the MSE is lower than that obtained in OLS, 1.2877, but, also in both cases, the value obtained for the number of conditions (although it has decreased) still indicates that the degree of approximate multicollinearity continues to be worrying.

Furthermore, the estimates obtained\footnote{Note that in the first case they are obtained from $\mathbf{K} = 0.0002705469 \cdot \mathbf{I}$  and in the second case from $\mathbf{K} = diag(0,0,0,0,0.0002705469)$.} re very similar, obtaining in both cases coefficients different from zero for \textbf{GNP}, \textbf{Unemployed} and \textbf{Armed Forces}. The goodness of fit, 0.9872, is very close to that of OLS, 0.9874.

On the other hand, if minimizing the condition number is prioritized, we obtain an MSE higher than that of OLS, although we obtain a condition number below 20, that is the threshold traditionally considered as worrying.

As in the previous cases, it is obtained that the coefficients associated with \textbf{GNP}, \textbf{Unemployed} and \textbf{Armed Forces} can be considered as significantly different from zero. The goodness of fit of 0.9838 and 0.9833 are also very close to that of OLS.

In summary, whether the priority is to minimize the MSE or the degree of multicollinearity, it can be concluded that the higher the gross national product, the greater the number of people employed, and the greater the number of unemployed and the number of people in the armed forces, the fewer the number of people employed. Therefore, applying the generalized ridge estimation has allowed us to establish more influential relationships between the variables considered than those initially obtained from the OLS estimation.

\subsection{Comparison with \textit{lmridge} and \textit{ridge} packages}

The estimates obtained for $k = 0, k_{HKB}, k_{HK}$ (options a) and b)) will be compared  below with those provided by the  \textit{lmridge} (\cite{lmridge}) and \textit{lrmest} (\cite{lrmest}) packages. he estimates for $k = k_{l,min}, k_{l}$ (options c) and d)) are not included as these packages do not have the option to estimate using the generalized ridge version.

Tables \ref{comparison_1} to \ref{comparison_3} show the estimate obtained for $k=0,k_{HKB}, k_{HK}$ as well as the values for the mean squared error (MSE), goodness of fit (GoF) and condition number (CN) if they are provided in the corresponding analysis. In parentheses are the confidence regions obtained using bootstrap methodology for the proposal made in this paper (called RR in the table) and the p-value of the individual significance provided by the \textit{lmridge} and \textit{ridge} packages. The estimates from the \textit{lmridge} and \textit{ridge} packages are obtained, respectively, with the options 'sc' and 'corrForm', which transform the variables into correlation form.

\begin{table}
    \centering
    \begin{tabular}{cccc}
        \hline
        $k=0$ & RR & \textit{lmridge} & \textit{ridge} \\
        \hline
        GNP deflator & -0.1489 (-1.0386, 0.9278) & -0.0485 (0.7081) & -0.048463 (0.7007) \\
        GNP & 2.0378 (-0.3464, 4.1346) & 0.072 (0.0365) & 0.072004 (0.0173) \\
        Unemployed & -0.1075 (-0.4182, 0.139) & -0.0040 (0.3548) & -0.004039 (0.3341) \\
        Armed Forces & -0.1111 (-0.2801, -0.0483) & -0.0056 (0.0636) & -0.005605 (0.0383) \\
        Population & -0.7992 (-2.3425, 0.8616) & -0.4035 (0.2264) & -0.403509 (0.2000) \\
        \hline
        MSE & 1.2877 & 238.2332 &  \\
        GoF & 0.9874 & 0.9874 &  \\
        CN & 61.5302 & 3785.966 &  \\
        \hline
    \end{tabular}
    \caption{Estimates for $k=0$ from the proposal made in this paper (called RR) and the \textit{lmridge} and \textit{ridge} packages} \label{comparison_1}
\end{table}

\begin{table}
    \centering
    \begin{tabular}{cccc}
        \hline
        $k=k_{HKB}=0.001186004$ & RR & \textit{lmridge} & \textit{ridge} \\
        \hline
        GNP deflator & 0.1789 (-0.3016, 0.7301) & 0.0582 (0.4963) & 0.058224 (0.48244) \\
        GNP & 1.1478 (0.2928, 1.7034) & 0.0406 (0.0138) & 0.040557 (0.00379) \\
        Unemployed & -0.2043 (-0.3853, -0.118) & -0.0077 (0.0172) & -0.007677 (0.00548) \\
        Armed Forces & -0.112 (-0.2593, -0.0517) & -0.0056 (0.0629) & -0.005653 (0.03974) \\
        Population & -0.1711 (-0.6806, 0.5708) & -0.0864 (0.6040) & -0.086396 (0.59401) \\
        \hline
        MSE & 1.6156 & 302.8489 &  \\
        GoF & 0.9861 & 0.9827 &  \\
        CN & 41.0823 & 1687.758 &  \\
        \hline
    \end{tabular}
    \caption{Estimates for $k=k_{HKB}=0.001186004$ from the proposal made in this paper (called RR) and the \textit{lmridge} and \textit{ridge} packages} \label{comparison_2}
\end{table}

\begin{table}
    \centering
    \begin{tabular}{cccc}
        \hline
        $k=k_{HK}=0.0002705469$ & RR & \textit{lmridge} & \textit{ridge} \\
        \hline
        GNP deflator & -0.0172 (-0.5988, 0.7884) & -0.0056 (0.9595) & -0.005582 (0.9586) \\
        GNP & 1.683 (0.0478, 2.8344) & 0.0595 (0.0287) & 0.059466 (0.0122) \\
        Unemployed & -0.1465 (-0.3874, -0.0096) & -0.0055 (0.1481) & -0.005505 (0.1205) \\
        Armed Forces & -0.118 (-0.2698, -0.0541) & -0.0056 (0.0608) & -0.005644 (0.0372) \\
        Population & -0.5494 (-1.5001, 0.6404) & -0.2774 (0.2941) & -0.277357 (0.2711) \\
        \hline
        MSE & 1.0189 & 190.1833 &  \\
        GoF & 0.9872 & 0.9855 &  \\
        CN & 54.3078 & 2949.343 &  \\
        \hline
    \end{tabular}
    \caption{Estimates for $k=k_{HK}=0.0002705469$ from the proposal made in this paper (called RR) and the \textit{lmridge} and \textit{ridge} packages} \label{comparison_3}
\end{table}
It is observed that:
\begin{itemize}
    \item The estimates obtained from \textit{lmridge} and \textit{ridge} simply differ in that the latter provide a greater number of decimal places. However, the associated p-value is different.
    \item The estimates obtained using RR are different to those obtained using \textit{lmridge} and \textit{ridge}.
    \item Both RR and \textit{lmridge} provide values for MSE, GoF and CN, although they are different from each other (the only coincidence observed is for GoF when $k=0$).
    \item The \textit{ridge} package does not provide information on MSE, GoF and CN.
    \item The \textit{lmridge} package identifies a smaller number of estimated coefficients as significantly different from zero than RR and \textit{ridge}.
\end{itemize}

The above comments suggest that there is a need to create a package in R that allows the calculation of regular and generalized ridge estimates, also providing estimates of the mean-squared error, goodness of fit, condition number and other measures for the detection of multicollinearity such as the variance inflation factor or coefficient of variation.

\section{Conclusions}
    \label{conclusion}

\cite{SalmeronGRR2021} deeply analyzes the generalized ridge estimator, paying special attention to the particular case in which the diagonal matrix $\mathbf{K}$ has a unique element nonnull $k_{l}$ and presenting its estimator, matrix of variances and covariances, trace, norm, mean squared error and goodness of fit. The analysis of the usefulness of the regression to mitigate multicollinearity is proposed as an interesting research line that is studied in this paper. The main conclusions are as follows:
\begin{itemize}
\item To treat multicollinearity with a ridge regression (in both regular and generalized cases), data should be transformed, and standardization is the most appropriate option. This conclusion is based on the previous papers presented by \cite{Garciaetal2016}, \cite{Salmeron2018CN} and \cite{Rodriguez2019R2,Rodriguezetal2021} that show that data should be standardized to obtain the desirable properties of continuity and monotony. The results obtained in this paper confirm this requirement for the case of the generalized ridge regression.
\item As a consequence of the previous point, the ridge regression is based on a noncentered augmented model, i.e., there is no intercept. This implies the following:
\begin{itemize}
\item Because there is no intercept, the model cannot present nonessential multicollinearity.
\item However, the multicollinearity could be nonessential multicollinearity defined in its generalized concept (two independent variables with low variability), although this paper analyzes the coefficient of variation of the regular and generalized ridge regressions and conclude that the degree of generalized nonessential multicollinearity will not be troubling.
\end{itemize}
\item Thus, essential multicollinearity is the only type of multicollinearity that could be troubling in this case. To detect it, we extend different measures (the coefficient of correlation, the variance inflation factor (VIF) and the condition number (CN)) to be applied in the regular and generalized ridge regressions and obtain different results:
\begin{itemize}
\item For the regular case, all the measures present adequate properties (continuity, monotony and minimum value), confirming that this option is appropriate to mitigate multicollinearity.
\item For the generalized case, the measures verify the conditions of continuity and minimum value, but the diagnostic measures increase as $k_{l}$ increases. However, from the analysis of the condition number, it is possible to conclude that a subset of values of $k_{l}$ that mitigate the multicollinearity can exit if the variable associated with the minimum eigenvalue is modified.
\end{itemize}
\item Finally, by following the previous indication, the adequate selection of $l$ can lead to values of $k_{l}$ that provide an estimator in which the degree of multicollinearity has been mitigated. The best behavior in terms of the MSE of the generalized ridge regression for these particular values can lead to estimators that not only mitigate the multicollinearity but also provide a MSE lower than that of the OLS.
\end{itemize}

Future lines of work include the creation of a package in R (\cite{RCoreTeam}) from the code available in Github that allows the application of the methodology presented to any user in the teaching, research and/or business environment.

%\section*{Acknowledgments}

%We would like to thank the reviewers for their valuable comments and suggestions.
%This work was supported by project PP2019-EI-02 of the University of Granada (Spain) and project A-SEJ-496-UGR20 I+D+i (FEDER Andaluc\'ia, 2014-2020).

\bibhang=1.7pc
\bibsep=2pt
\fontsize{9}{14pt plus.8pt minus .6pt}\selectfont
\renewcommand\bibname{\large \bf References}
%\begin{thebibliography}{11}
\expandafter\ifx\csname
natexlab\endcsname\relax\def\natexlab#1{#1}\fi
\expandafter\ifx\csname url\endcsname\relax
  \def\url#1{\texttt{#1}}\fi
\expandafter\ifx\csname urlprefix\endcsname\relax\def\urlprefix{URL}\fi

%% use bibfile
\bibliographystyle{chicago}      % Chicago style, author-year citations
\bibliography{bib}

%%%%%%%%%%%%%%%%%%%%%%%%%%%%%%%%%%%%%%%%%%%%%%%%%%%%%%%%%%%%%%%%%%%%%%%%%%%%%%%%%%%%%%%%%%%%%%%%%%%%%%%%%%%%%%%%%%%%%%%%%%%%
\vskip .65cm
\noindent
Rom\'an Salmer\'on G\'omez
\vskip 2pt
\noindent
E-mail: (romansg@ugr.es)
\vskip 2pt

\noindent
Catalina B. Garc\'ia Garc\'ia
\vskip 2pt
\noindent
E-mail: (cbgarcia@ugr.es)

\vskip 2pt

\noindent
Guillermo Hortal Reina
\vskip 2pt
\noindent
E-mail: (ghorrei@correo.ugr.es)

% \vskip .3cm
%\centerline{(Received ???? 20??; accepted ???? 20??)}\par
\end{document}